\documentclass[twocolumn]{aastex63}
\usepackage{xcolor}
\usepackage{hyperref}
\usepackage{enumerate}
\usepackage [caption=false]{subfig}
\usepackage{natbib}
\usepackage{url}
\usepackage{amsthm,amsmath,amssymb}

\usepackage{mathrsfs}
\newcommand\Nu{{\it NuSTAR} }
\newcommand\XMM{{\it XMM-Newton} }

\newcommand\ec{$E_{\rm cut} $ }

\begin{document}

\title{
On joint analysing {\it XMM-NuSTAR} spectra of active galactic nuclei
}
\correspondingauthor{Jia-Lai Kang \& Jun-Xian Wang} \email{ericofk@mail.ustc.edu.cn, jxw@ustc.edu.cn}

\author{Jia-Lai Kang} 
\author{Jun-Xian Wang}
\affiliation{CAS Key Laboratory for Research in Galaxies and Cosmology, Department of Astronomy, University of Science and Technology of
China, Hefei, Anhui 230026, China; ericofk@mail.ustc.edu.cn, jxw@ustc.edu.cn}
\affiliation{School of Astronomy and Space Science, University of Science and Technology of China, Hefei 230026, China}

\begin{abstract}
A recently released \XMM technical note has revealed a significant calibration issue between \Nu and \XMM EPIC, and provided an empirical correction to EPIC effective area. To quantify the bias caused by the calibration issue to joint analysis of {\it XMM-NuSTAR} spectra and verify the effectiveness of the correction, in this work we perform joint-fitting of \Nu and EPIC-pn spectra for a large sample of 104 observation pairs of 44 X-ray bright AGN. 
The spectra were extracted after requiring perfect simultaneity between \XMM and \Nu exposures (merging GTIs from two missions) to avoid bias due to rapid spectral variability of AGN.
Before the correction, the EPIC-pn spectra are systematically harder than corresponding \Nu spectra by $\Delta \Gamma \sim 0.1$, subsequently yielding significantly underestimated cutoff energy \ec and the strength of reflection component $R$ when performing joint-fitting. We confirm the correction is highly effective and can commendably erase the discrepancy in best-fit $\Gamma$, \ec and $R$, and thus we urge the community to apply the correction when joint-fitting {\rm XMM-NuSTAR} spectra. Besides, we show that as merging GTIs from two missions would cause severe loss of \Nu net exposure time, in many cases joint-fitting yields no advantage compared with utilizing \Nu data alone. 
We finally present a technical note on filtering periods of high background flares for \XMM EPIC-pn exposures in the Small Window mode.
\end{abstract}

\keywords{Galaxies: active – Galaxies: nuclei  – X-rays: galaxies }

\section{Introduction} \label{sec:intro}
\par The Nuclear Spectroscopic Telescope Array \citep[{\it NuSTAR}; ][]{Harrison_2013}, the first direct-imaging hard X-ray mission with a spectral coverage from 3 to 79 keV, has remarkably boosted the study of various high-energy phenomenons.
Active galactic nuclei are the dominant population in the extragalactic X-ray sky. According to the widely accepted disk-corona paradigm, the primary X-ray emission of AGN originates in a hot and compact region, namely the corona \citep{Haardt_1991, Haardt_1993}. The seed photons from the accretion disk are up-scattered to the X-ray band through inverse Comptonization, producing the observed power-law continuum, with a cutoff in the high energy end, which is a direct indicator of the coronal temperature $T_{\rm e}$. 
For AGN studies, \Nu opens a new window to detect/measure the X-ray high energy cutoff $E_{cut}$ (and effectively, the corona temperature $T_e$) in many sources \citep[e.g., ][]{Brenneman_2014, Matt_2015, Fabian_2015, Kamraj_2018, Balokovi_2020, Kang_2022}. Meanwhile, \Nu observations are also essentially helpful to constrain the reflection component in the X-ray spectra \citep[e.g., ][]{Parker_2014, Kara_2015, Wilkins_2015, Panagiotou_2019}.  

In practice, \Nu data are often analysed in association with coordinated data of other X-ray missions, to achieve a broader energy coverage and higher spectral signal to noise ratio at $<$ 10 keV, which could supposedly improve the constraints to X-ray spectral parameters. 
Among the coordinated missions, \XMM \citep{Jansen_2001_XMM} is of most significance, with the longest coordinated observing time since \Nu Cycle 2\footnote{\url{https://heasarc.gsfc.nasa.gov/docs/nustar/nustar_prop.html}}. Thanks to the large effective areas of the EPIC cameras, \XMM could provide high-quality spectra in 0.3 -- 10 keV, as a nice complement to {\it NuSTAR}. 

However, proper inter-instrument calibration is criticial prior to analyzing joint observations from different observatories. 
A possible calibration issue between \Nu and \XMM had been reported in literature \citep[e.g., ][]{gamma_xmm_1, gamma_xmm_3, gamma_xmm_2}.
Such calibration issues could be the fundamental cause of the descrepancy in the measurements of $E_{\rm cut}$ or $T_{\rm e}$ in individual AGN, between studies fitting \Nu spectra alone \citep[e.g., ][]{Kamraj_2018, Rani_2019, Kang_2020, Panagiotou_2020, Akylas_2021, Kang_2022, Pal_2023}, and those joint-fitting quasi-simultaneous data from other missions \citep[e.g., ][]{Marinucci_2014, Tortosa_2018, Zhangjx2018, Molina_2019, Balokovi_2020, Hinkle_2021, Kamraj_2022, Pal_2022}. 

The calibration issue between \Nu and \XMM
was confirmed by a recently released \XMM calibration technical note\footnote{\url{https://xmmweb.esac.esa.int/docs/documents/CAL-TN-0230-1-3.pdf}\label{footnote2}}, in which the EPIC-pn spectra of the Crab Nebula were found to be systematically harder than those of {\it NuSTAR}, with $\Gamma^{\rm pn} - \Gamma^{\rm NuSTAR} \sim -0.1$. An empirical correction of the effective area is then implemented (after SAS 20.0) in the newest \XMM calibration files (but not deployed by default in the pipeline), improving the cross-calibration with {\it NuSTAR}.
Such a correction however has not been adopted by most joint-analyses of {\it XMM-NuSTAR} spectra of AGN in literature. It is both technically and scientifically important to investigate how this calibration issue biases the X-ray spectral measurements of AGN, verify the effectiveness of the correction, and if yes, provide the updated X-ray spectral parameters after applying the calibration correction for AGN with joint {\it XMM-NuSTAR} observations.

\section{Sample and Data Reduction} \label{sec:data}

\par We match the 817 Seyfert galaxies in the 105 month {\it Swift}/BAT catalog \citep{Oh_2018} with archival \Nu and \XMM observations (as of 2023 August). In this work we focus on the EPIC-pn data of \XMM \citep{Struder_2001_PN}, which are most widely used in literature. However, the empirical correction of the effective areas for \XMM EPIC-MOS \citep{Turner_2001} is exactly the same (see the calibration note in footnote \ref{footnote2}), so the conclusions about the cross-calibration should also apply to the MOS data. We search for simultaneous observation pairs of the two missions, requiring an overlapping exposure time $>$ 5 ks between \Nu and EPIC-pn, obtaining 176 observation pairs of 94 sources. Note one \Nu exposure may overlap in time with more than one \XMM exposures, and vice versa. 
Observation pairs with too few counts, or complicated spectra badly fitted with the model in \citet{Kang_2022}, are further dropped, resulting in a final sample of 104 observation pairs (consisting of 100 \Nu and 102 \XMM observations) of 44 sources (see Table \ref{tab:obs})\footnote{21 observation pairs in this work were also adopted in the aforementioned technical note, but only used to show the improvement of the statistics (reduction of $\chi^2$), irrelevant to the derivation of the empirical correction of ARF. We confirm the inclusion of these observations in this work does not bias our analysis.}. 

\par
\begin{figure}
\centering
\subfloat{\includegraphics[width=0.4\textwidth]{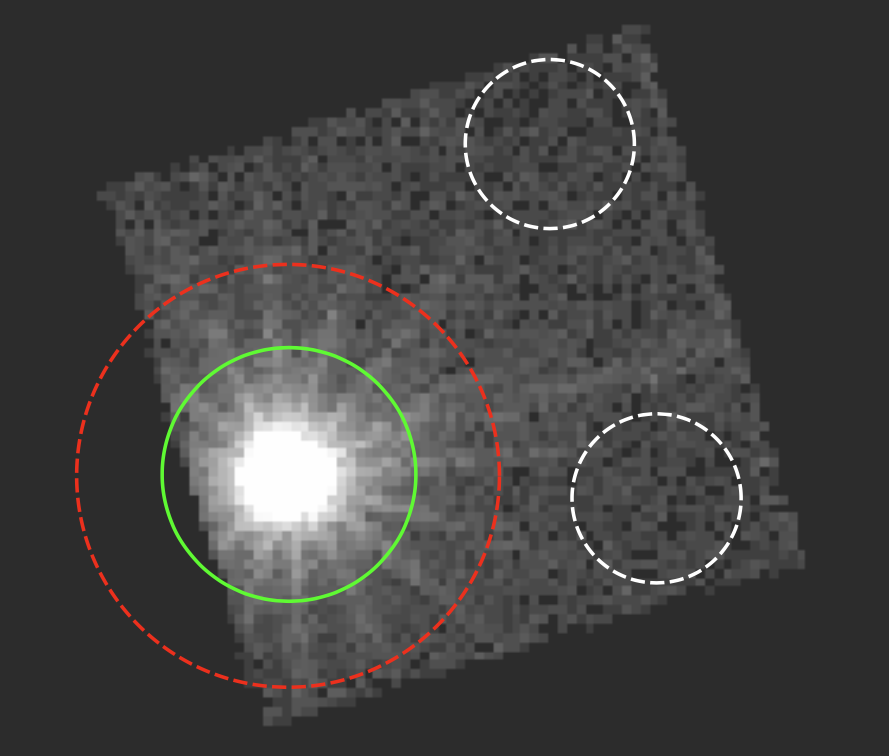} }
\caption{\label{fig:region}An example (\XMM obsid 0741330101) of the adopted regions when processing the \XMM EPIC-pn data in Small Window mode. The green circle is the source region, optimally determined by \textit{eregionanalyse}, while the two white circles with a radius of 40\arcsec\ are the background regions. Outside the red circle with a radius of 100\arcsec\ is the source-free region, used to filter intervals with flaring background. }
\end{figure}

 Raw \Nu data are reduced with \textit{nupipeline}, part of the HEASoft package (version 6.32.1), with calibration files of version 20210824. We extract the \Nu spectra using \textit{nuproducts}, adopting a circular source region with a radius of 60$\arcsec$ for each source, and an annulus from 120$\arcsec$ to 150$\arcsec$ for background extraction. 
Raw \XMM data are processed with the Science Analysis Software (SAS 20.0.0) and current calibration files (CCF 3.13). We extract the source spectra from a circular region, the radius of which is optimally determined by \textit{eregionanalyse}, with background from nearby source-free regions (see Figure \ref{fig:region} for example). \textit{epatplot} is employed to confirm the pile-up effect is negligible in all the observations. Two ancillary response files (ARF) are created for each observation, with/without applying the aforementioned empirical correction of the effective area, by setting the parameter $applyabsfluxcorr$ of the task \textit{arfgen}. Finally, all the spectra are rebinned to attain a minimum of 50 counts per bin using \textit{grppha}. 

\par In the above paragraph we describe a typical process of the data reduction, but omit an ambiguous step of reducing the \XMM data, i.e., filtering periods suffering from flaring background. Solar protons could produce flaring EPIC background, which is highly unpredictable and affects a large fraction of \XMM observation time \citep{Read_2003, Carter_2007}. As recommended by the user guide\footnote{\url{https://xmm-tools.cosmos.esa.int/external/xmm_user_support/documentation/sas_usg/USG/epicbkgfiltering.html}}, a viable method is to extract a rate curve (usually with a time bin $<=$ 100 s) in the source-free regions, of only valid single events with energy between 10 keV and 12 keV (``PATTERN==0 \&\& PI IN [10000:12000]" in the expression of \textit{evselect}). 
A new Good Time Interval (GTI) file will then be generated, dropping the periods with count rate above certain threshold, a recommended value of which is 0.4 counts/s for the EPIC-pn camera in the Full Frame mode. 

\par
\begin{figure}
\centering
\subfloat{\includegraphics[width=0.48\textwidth]{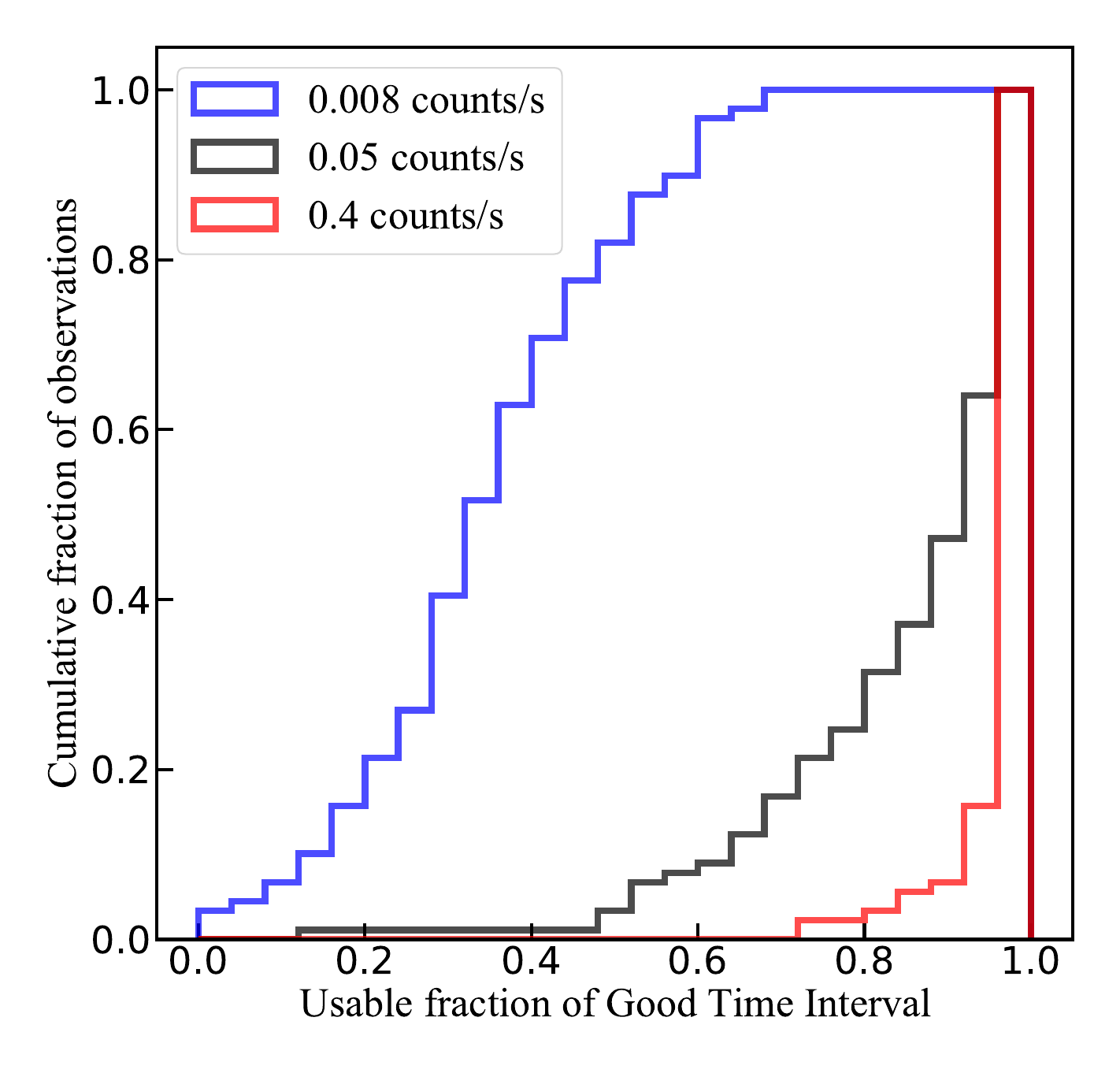} }
\caption{\label{fig:gti} Usable fraction of the Good Time Interval after filtering the periods showing flaring background with different thresholds, for the 85 \XMM exposures in SW mode. }
\end{figure}

 However, as shown in Table \ref{tab:obs}, most observations (85/102) of these bright AGN are operated in the Small Window mode (hereafter SW mode) to avoid serious pile-up effect, which has a much smaller FOV (1/36) than the Full Frame mode (hereafter FF mode). The value of the count rate threshold is hence supposed to be smaller for the SW mode, while the time bin of the rate curve should be large enough to contain sufficient counts. Therefore, for these SW observations, we try three different thresholds: 1) 0.4 counts/s, same as the FF mode; 2) 0.008 counts/s, scaled down because in the SW mode a much smaller sky area (see Figure \ref{fig:region}) could be used to extract 10 -- 12 keV light curve for high background filtering; and 3) 0.05 counts/s, a moderate threshold between 1) and 2); and adopt a time bin of 500 s. The consequent usable fraction of GTI is shown in Figure \ref{fig:gti}. Apparently, the threshold of 0.008 counts/s is too strict, leading to less than 40\% GTI usable in more than half observations. 
Meanwhile, the threshold of 0.4 counts/s is likely too loose, causing insufficient filtering, the bias of which we shall show in \S \ref{sec:filter}. On the other hand, the threshold of 0.05 counts/s seems to be reasonable, and the corresponding distribution of usable GTI is similar to the one in \citet{Read_2003}, of a detailed study of EPIC background with FF observations. 

\par
\begin{figure}
\centering
\subfloat{\includegraphics[width=0.53\textwidth]{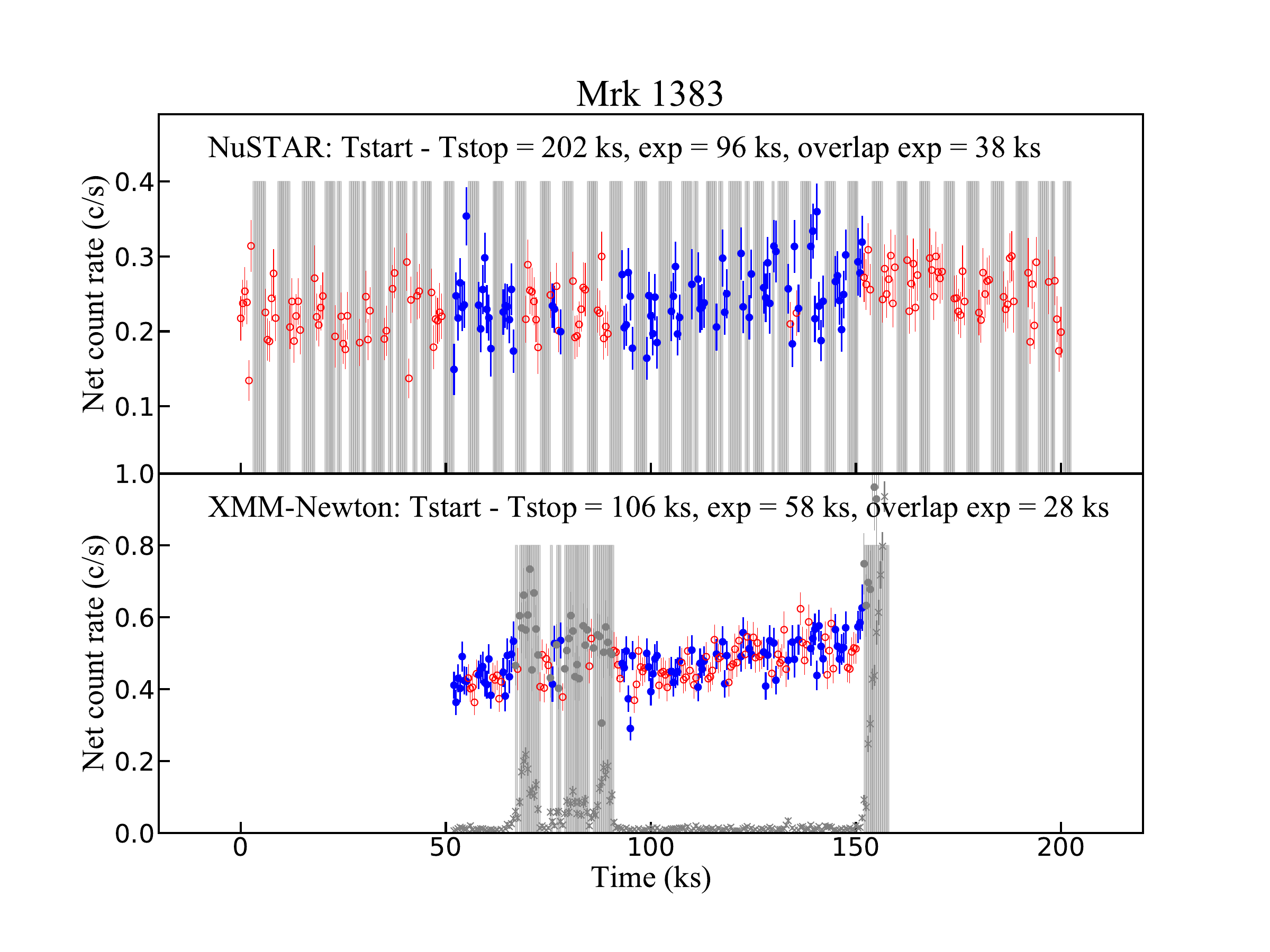}}
\caption{\label{fig:Mrk1383} Example \Nu FPMA 3 -- 78 keV (ObsID: 60501049002) and \XMM EPIC-pn 3 -- 10 keV (ObsID: 0852210101) light curves (with a time bin of 500 s) of Mrk 1383, to illustrate 
the merge of \Nu and EPIC-pn GTIs. The grey shades represent the dropped time intervals for each instrument, mainly due to the Earth occultation for \textit{NuSTAR}, and flaring background for EPIC-pn. The blue dots represent the remaining data after merging the GTIs,
while the red open circles mark the data dropped due to the merging of GTI. For EPIC-pn (lower panel), we over-plot the background rate curve (grey stars) used to filter the intervals with flaring background, and the corresponding data points filtered out from the source light curve (grey circles). 
The duration of the exposures and net exposure time (before/after merging GTIs) are labelled, respectively. 
 }
\end{figure}

 Therefore, in our analyses below we adopt a threshold of 0.05 counts/s (unless otherwise stated, see the lower panel of Figure \ref{fig:Mrk1383} for an example) to filter the flaring background for SW observations\footnote{We adopt a threshold of 0.4 counts/s and 0.2 counts/s, for FF and LW observations respectively. The sample of them is too small to perform similar analysis.}, while in \S \ref{sec:filter}, using \Nu spectra as reference, we shall show that such a threshold is indeed appropriate, whereas a looser threshold (0.4 counts/s) will lead to 
insufficient background subtraction and hence biased spectral parameters. After high background screening, the usable GTI of \Nu and EPIC-pn are merged to find the overlapped part (see Figure \ref{fig:Mrk1383} for an example, and Table \ref{tab:sole} for the net exposure time before/after the GTI merge for each \Nu exposure). The spectra of \Nu and EPIC-pn for a certain observation pair are then extracted from the overlapped GTI (i.e., with perfect simultaneity). This step is necessary to avoid potential bias caused by rapid X-ray spectral variability of AGN \citep[e.g.][]{Wu_2020}, that two spectra without perfect simultaneity may have intrinsically different spectral shape. 

\section{Spectral fitting} \label{sec:fitting}
\par We perform spectral fitting using XSPEC \citep{Arnaud_1996} and the $\chi^2$ statistics. All the errors and the upper/lower limits of the parameters are derived following the $\Delta \chi^2 = 2.71$ criteria (corresponding to the 90\% confidence level for one interesting parameter). Spectral fitting is carried out in the 3--78 keV band for \Nu and 3--10 keV band for EPIC-pn. We drop the data $<$ 3 keV of EPIC-pn because 1) to avoid the influence of other spectral components, such as soft X-ray excess and complicated absorption features, in the soft band, and 2) the aformentioned empirical ARF correction for EPIC-pn is only available between 3 -- 12 keV. 
For each \Nu observation, the spectra of FPMA and FPMB are jointly fitted with a cross-normalization \citep{Madsen_2015}.

\par Following \citet{Kang_2022}, we employ the model $ zphabs \times (pexrav + zgauss)$ to fit the spectra. Among them, $zphabs$ models the intrinsic photoelectric absorption, $pexrav$ \citep{Magdziarz_1995} models an exponentially cut-off power law with a neutral reflection component, and $zgauss$ models the Fe K$\alpha$ line. 
During the fitting, the absorption column density $N_{\rm H}$, photon index $\Gamma$, high energy cutoff $E_{\rm cut}$, reflection strength $R$ are set free to vary. We deal with the Fe K$\alpha$ line in the same way as \citet{Kang_2022}. We firstly fix the line at 6.4 keV in the rest frame and the line width at 19 eV \citep[the mean Fe K$\alpha$ line width in AGNs measured with Chandra HETG;][]{Shu_2010} to model a neutral narrow Fe K$\alpha$ line. Then we set the line width free to vary and adopt the corresponding results if such a variable line width prominently improves the fitting ($\Delta \chi^2 > 5$). To simultaneously show the discrepancy between two missions and how the measurement of \ec and $R$ is biased, we conduct both independent and joint fit of the \Nu and EPIC-pn spectra. We firstly fit the \Nu spectra alone and obtain the best-fit parameters ($\Gamma^{\rm Nu}$, $E_{\rm cut}^{\rm Nu}$ and $R^{\rm Nu}$ in Table \ref{tab:obs}). Then we fit the corresponding EPIC-pn spectrum, with \ec and R fixed at the best-fit results of \Nu spectra, as both parameters are barely constrained by EPIC-pn data. We then derive $\Gamma^{\rm pn}$ and $\Gamma^{\rm pn-Cor}$, for effective area uncorrected/corrected EPIC-pn data respectively, to highlight the effect of the calibration issue and the correction. Finally, we jointly fit the \Nu and EPIC-pn spectra to measure the $E_{\rm cut}^{\rm joint}$, $R^{\rm joint}$, $E_{\rm cut}^{\rm joint-Cor}$, and $R^{\rm joint-Cor}$. During the joint-fitting, all the parameters of the EPIC-pn spectrum are linked with \Nu spectra, except a variable constant to account for the absolute normalization between the two missions.

\section{Discussion} \label{sec:discussion}
\subsection{The calibration between \Nu and \XMM}

\begin{figure*}
\centering
\subfloat{\includegraphics[width=0.33\textwidth]{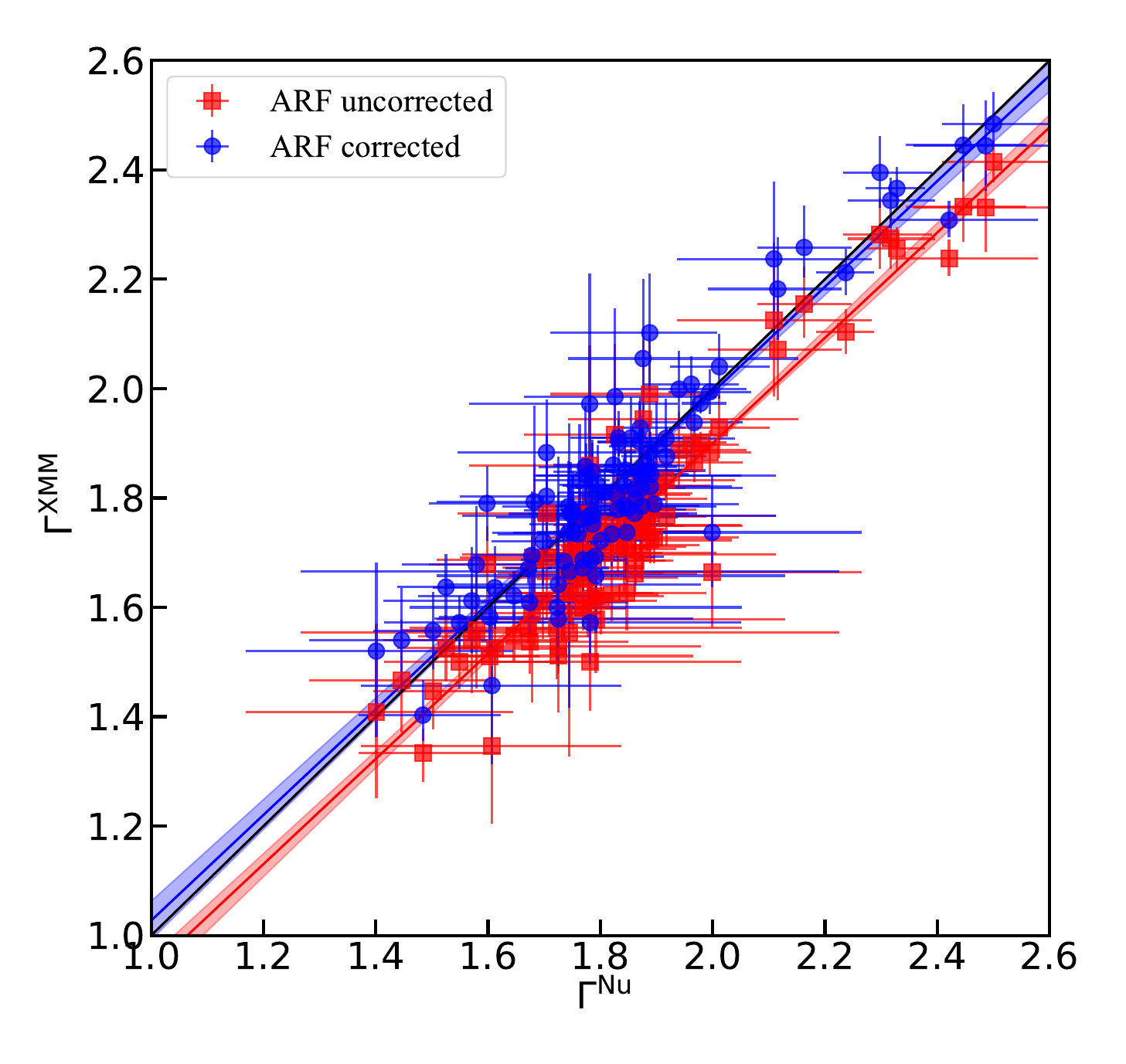} }
\subfloat{\includegraphics[width=0.33\textwidth]{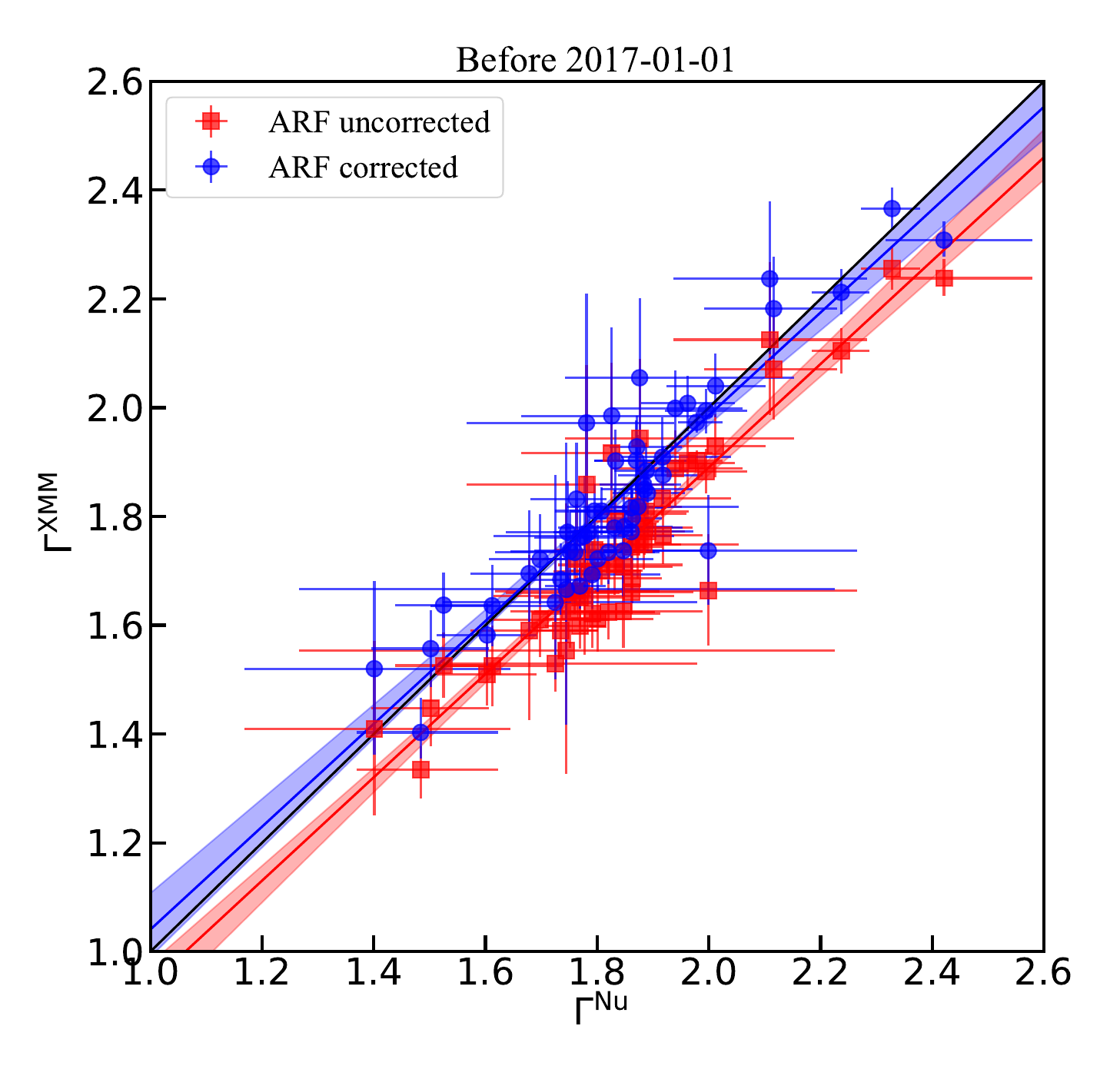} }
\subfloat{\includegraphics[width=0.33\textwidth]{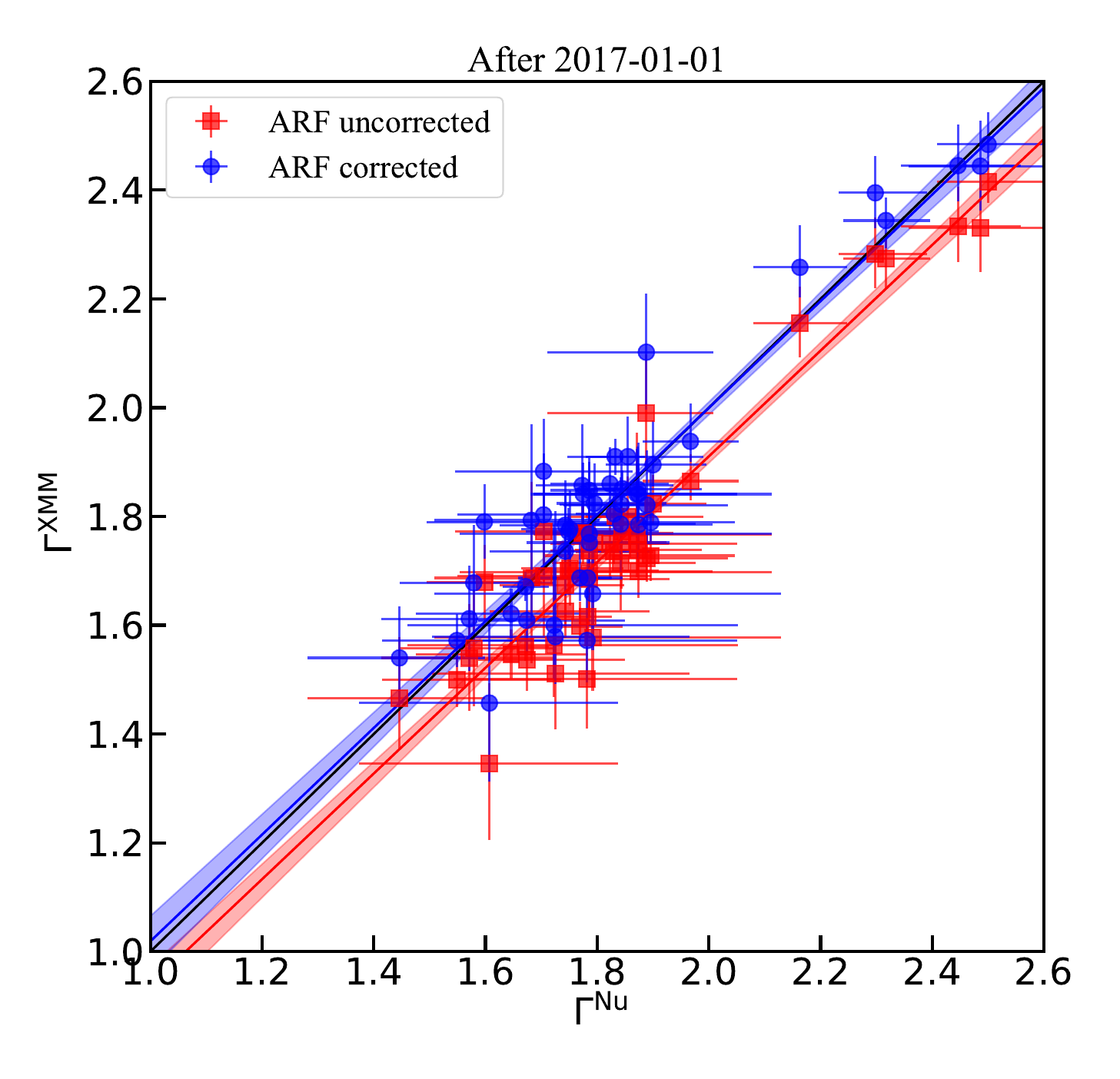} }
\caption{\label{fig:gamma} The left panel: photon index $\Gamma$ of the \Nu spectra versus those of the EPIC-pn spectra before (red square) or after (blue circle) correcting the effective area (corresponding to $\Gamma^{\rm Nu}$, $\Gamma^{\rm pn}$ and $\Gamma^{\rm pn-Cor}$ in Table \ref{tab:obs} respectively). 
The colored solid lines show the linear regression results (in comparison with the black 1:1 line), with the shadows showing the 1$\sigma$ uncertainty derived through bootstrapping the sample.
The middle and right panels show the case for observations before/after 2017-01-01. }
\end{figure*}

\begin{figure*}
\centering
\subfloat{\includegraphics[width=0.45\textwidth]{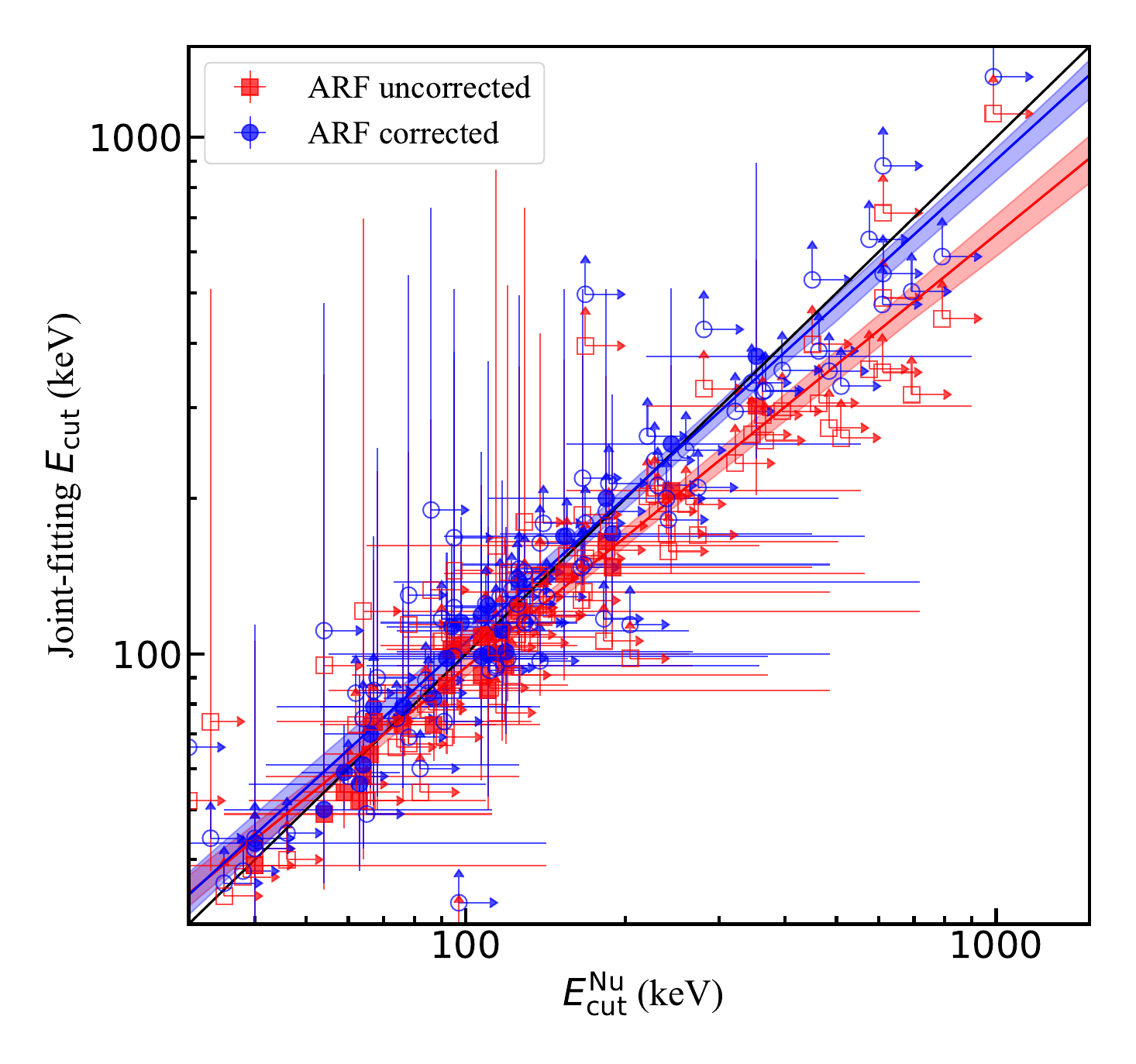} }
\subfloat{\includegraphics[width=0.45\textwidth]{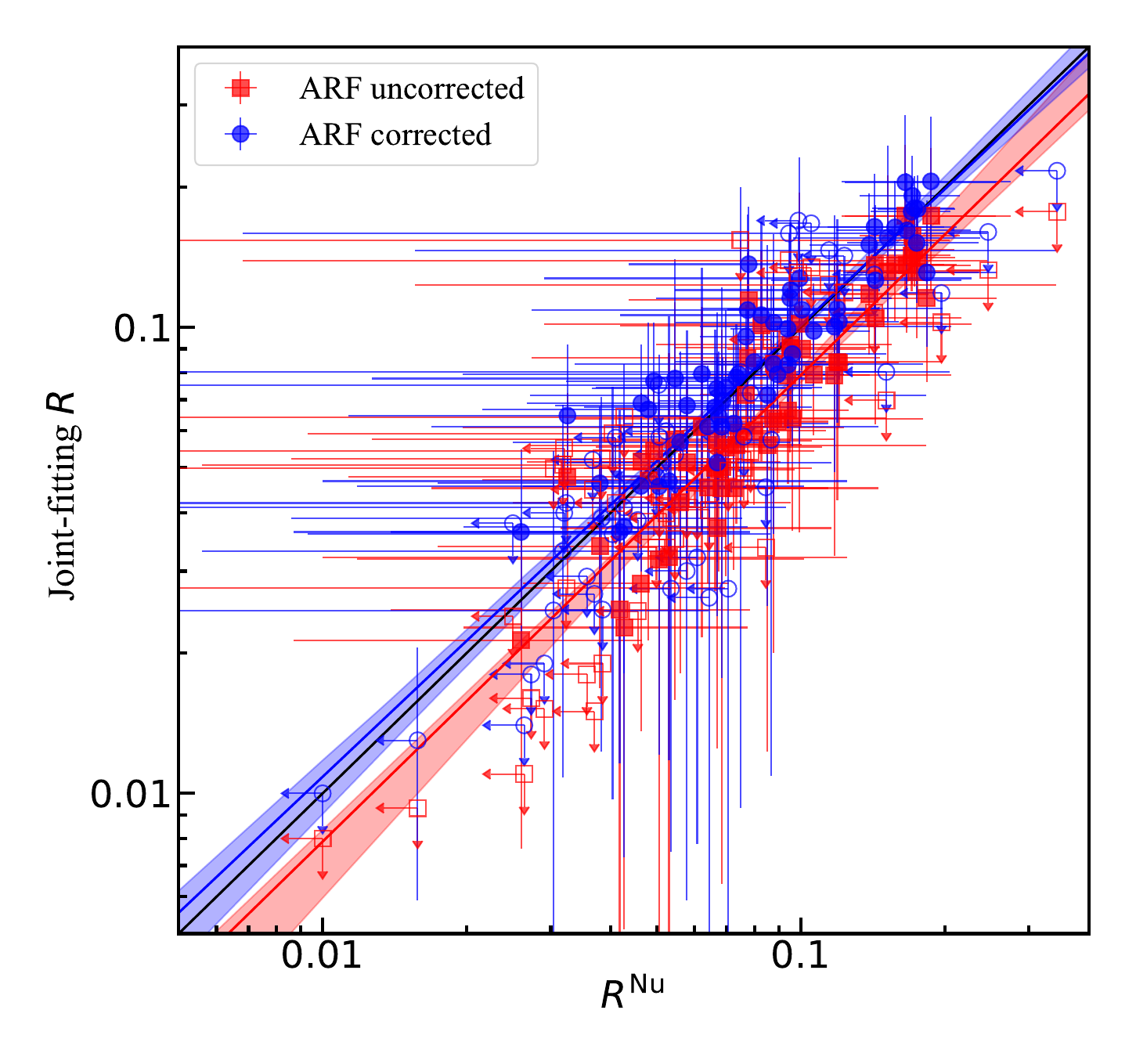} }
\caption{\label{fig:result} \ec and $R$ derived through fitting the \Nu spectra alone versus joint-fitting with EPIC-pn spectra. We perform linear regression in logarithmic space  
with \textit{asurv} \citep{Feigelson_1985} to handle the censored data points (as hollow markers). The colored solid lines show the linear regression results (in comparison with the black 1:1 line), with the shadow showing the 1$\sigma$ uncertainty derived through bootstrapping the sample. }
\end{figure*}

\par We show the best-fit $\Gamma$ of \Nu and EPIC-pn spectra in Figure \ref{fig:gamma}. Patently, the EPIC-pn spectra without the correction are systematically and significantly harder than the coordinated \Nu spectra. Specifically, we find that mean $\Gamma^{\rm Nu} = 1.84 \pm 0.02$, while mean $\Gamma^{\rm pn} = 1.74 \pm 0.02$. Meanwhile, the empirical correction of the effective area seems able to completely erase the discrepancy, resulting in mean $\Gamma^{\rm pn-Cor} = 1.84 \pm 0.02$. Furthermore, the two subsamples of observations before/after 2017-01-01 {(roughly equally divided)} show that the discrepancy does not evolve with time. 

\par In Figure \ref{fig:result} we compare the joint-fitting derived \ec and $R$ (before and after applying the correction respectively) with those derived through fitting \Nu spectra alone.
Apparently, the smaller $\Gamma^{\rm pn}$ also leads to smaller \ec and $R$ when perform joint-fitting, due to the strong positive degeneracy between $\Gamma$ and \ec as well as between $\Gamma$ and $R$ \citep[e.g.,][]{Molina_2019, Panagiotou_2020, Kang_2021, Kang_2023}. Meanwhile, the correction is also highly effective for these two parameters. 

\par Therefore, without the correction, the calibration issue between \Nu and EPIC-pn will bias the measured \ec and $R$ towards lower values.
We note the exact strength of the bias would depend on the statistical significance of the EPIC-pn data. Note in this work we require a perfect simultaneity between \Nu and  EPIC-pn data to avoid additional bias due to intrinsic spectral variation, which is not the case in most works. Neglecting the intrinsic variability, the effect of the bias would be stronger if the EPIC-pn exposure is longer than the \Nu one, as relatively longer EPIC-pn exposure could enhance the dominance of the uncorrected EPIC-pn spectrum's contribution to the joint fitting. Besides, for sources at high redshift or with high accretion rate and very steep spectra \citep[e.g.,][]{Lanzuisi_2016, Lanzuisi_2019, Tortosa_2021}, the bias is also expected to be more serious as the EPIC-pn data could be more dominant during the spectral fitting in these cases. We therefore urge the community to adopt the \XMM ARF correction when joint-fitting {\it XMM-NuSTAR} spectra spectra (but note the correction at $<$ 3 keV is yet unavailable).

\par Moreover, although in this work we focus on how the measurements of $\Gamma$, \ec and $R$ are biased, the calibration issue could also bias the measurements of other parameters derived though joint fitting, for example, the spin of the blackhole $a$ \citep[e.g.,][]{Risaliti_2013, Porquet_2019, Jiang_2022} and the absorption column density $N_{\rm H}$ in Compton-thick sources \citep[e.g.,][]{Marchesi_2022, Silver_2022, Sengupta_2023}. Compared with \ec and $R$, deriving these parameters is more model-dependent and the effect of the bias should be inspected case by case (which is beyond the scope of this work). 

\subsection{Filter the flaring background of \XMM in Small Window mode} \label{sec:filter}

\par
\begin{figure}
\centering
\subfloat{\includegraphics[width=0.45\textwidth]{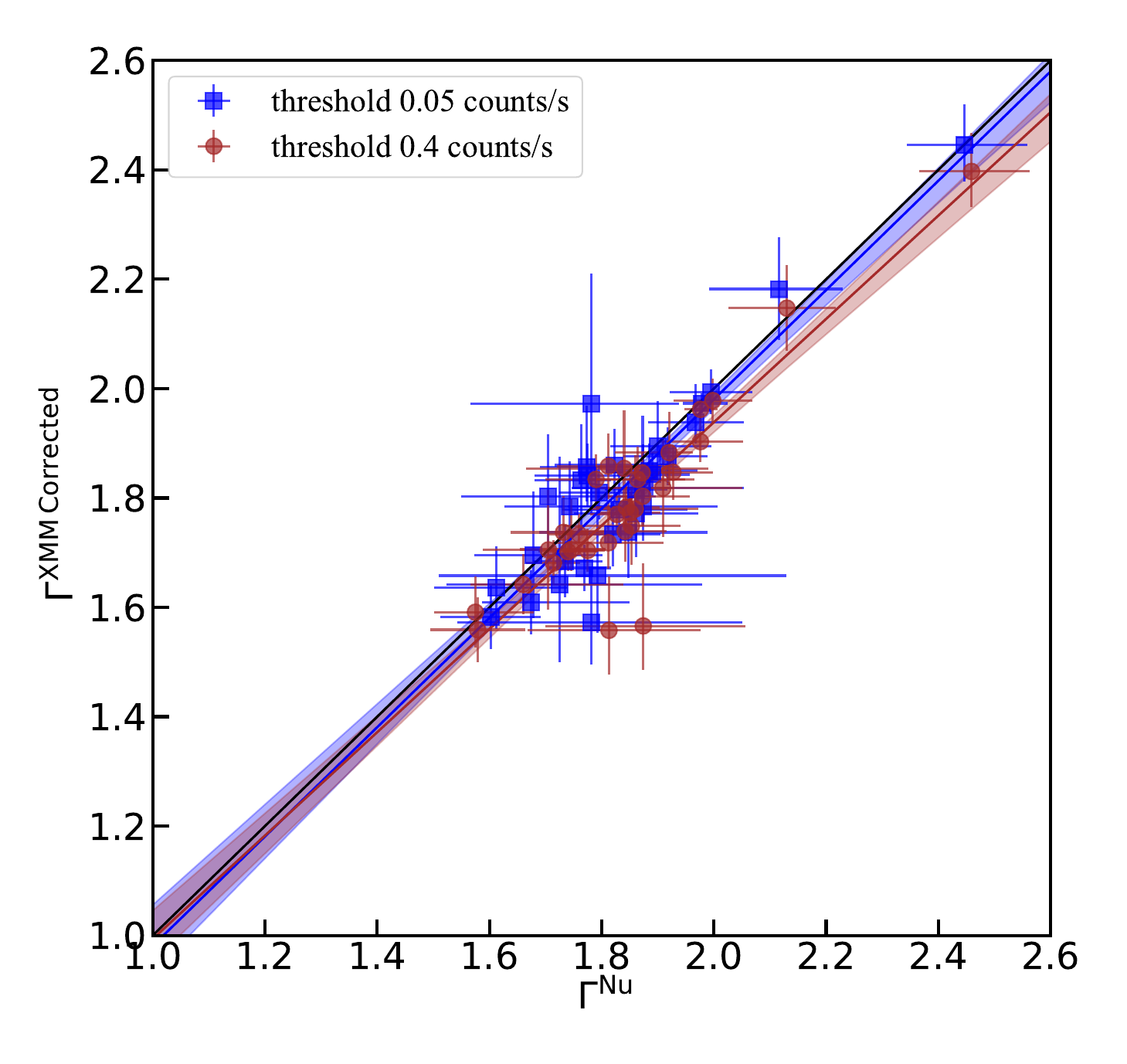} }
\caption{\label{fig:filter}The photon index $\Gamma$ (y-axis) derived using the effective area corrected \XMM EPIC-pn spectra, after filtering the periods showing flaring background with a threshold of 0.05 counts/s (blue squares) or 0.4 counts/s (brown circles), versus those derived using \Nu spectra (x-axis), for the 33 observation pairs with significant flaring background. The colored solid lines show the linear regression results (in comparison with the black 1:1 line), with the shadow showing the 1$\sigma$ uncertainty derived through bootstrapping the sample. }
\end{figure}

 Using \Nu spectra as references for ARF corrected \XMM EPIC-pn spectra, we investigate how the choice of the threshold count rate to filter high background flares affects the spectral fitting results. For the \XMM SW mode observations, we now take a threshold of 0.4 counts/s to filter high background intervals, regenerate the common GTI, and in which we extract the corresponding \Nu and EPIC-pn spectra. To highlight the impact of the flaring background, we limit the comparison to the observations with significant background flaring, through requiring that
in $>$ 10\% of all usable GTI, the 10 -- 12 keV count rate of the light curve we generated to filter background flares lies between 0.05 and 0.4 counts/s. 
33 of the 89 SW observations pairs meet this criterion, the $\Gamma^{\rm Nu}$ versus $\Gamma^{\rm pn-Cor}$ of which are shown in Figure \ref{fig:filter}. 

\par After applying the \XMM released ARF correction, the \XMM EPIC-pn spectra generated with a filtering threshold of 0.4 counts/s are still slightly harder than the \Nu ones (see Figure \ref{fig:filter}), indicating the fitting is biased. We interpret the bias as an underestimation of the flaring (soft protons induced) background,
as their spectra are apparently much harder than those of AGN \citep[see \S 4 in][]{Kuntz_2008}, the underestimation of which will hence lead to a harder net spectrum. For SW observations, the target object is always placed near the aim point of the telescope, where the flaring background will be stronger than that in the region used for background subtraction, due to the vignetting effect of the background flares \citep[see Figure 17 in][]{Kuntz_2008}. Our finding shows that despite the small FOV of the SW mode, the spatial unevenness of the background is non-negligible when the flaring background is strong and not properly filtered, which will bias the spectral fitting even for these bright AGN. 

\par On the other hand, Figure \ref{fig:filter} also confirms that a threshold of 0.05 counts/s is appropriate to filter high background flares in the Small Window mode of \XMM EPIC-pn observations. After filtering background flares, the residual quiescent soft protons induced background, along with other potential components \citep[e.g.,][]{DeLuca_2004}, cause no significant bias at least for these bright AGN.

\subsection{Do simultaneous \XMM data really help?} 

\begin{figure}
\centering
\subfloat{\includegraphics[width=0.45\textwidth]{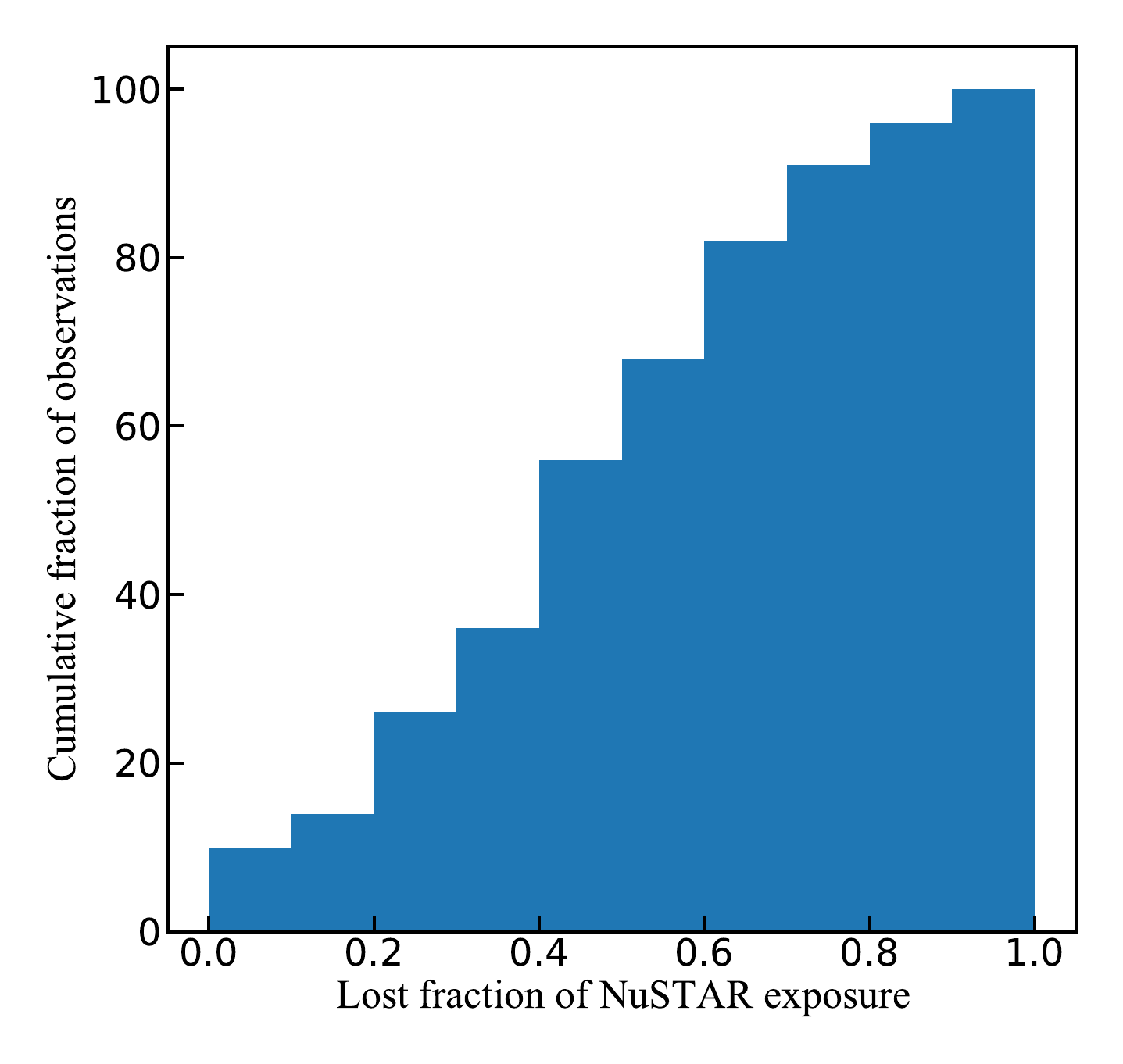}}
\caption{\label{fig:explost}Lost fraction of the \Nu net exposure after requiring a perfect simultaneity between \Nu and \XMM data. }
\end{figure}

\begin{figure*}
\centering
\subfloat{\includegraphics[width=0.33\textwidth]{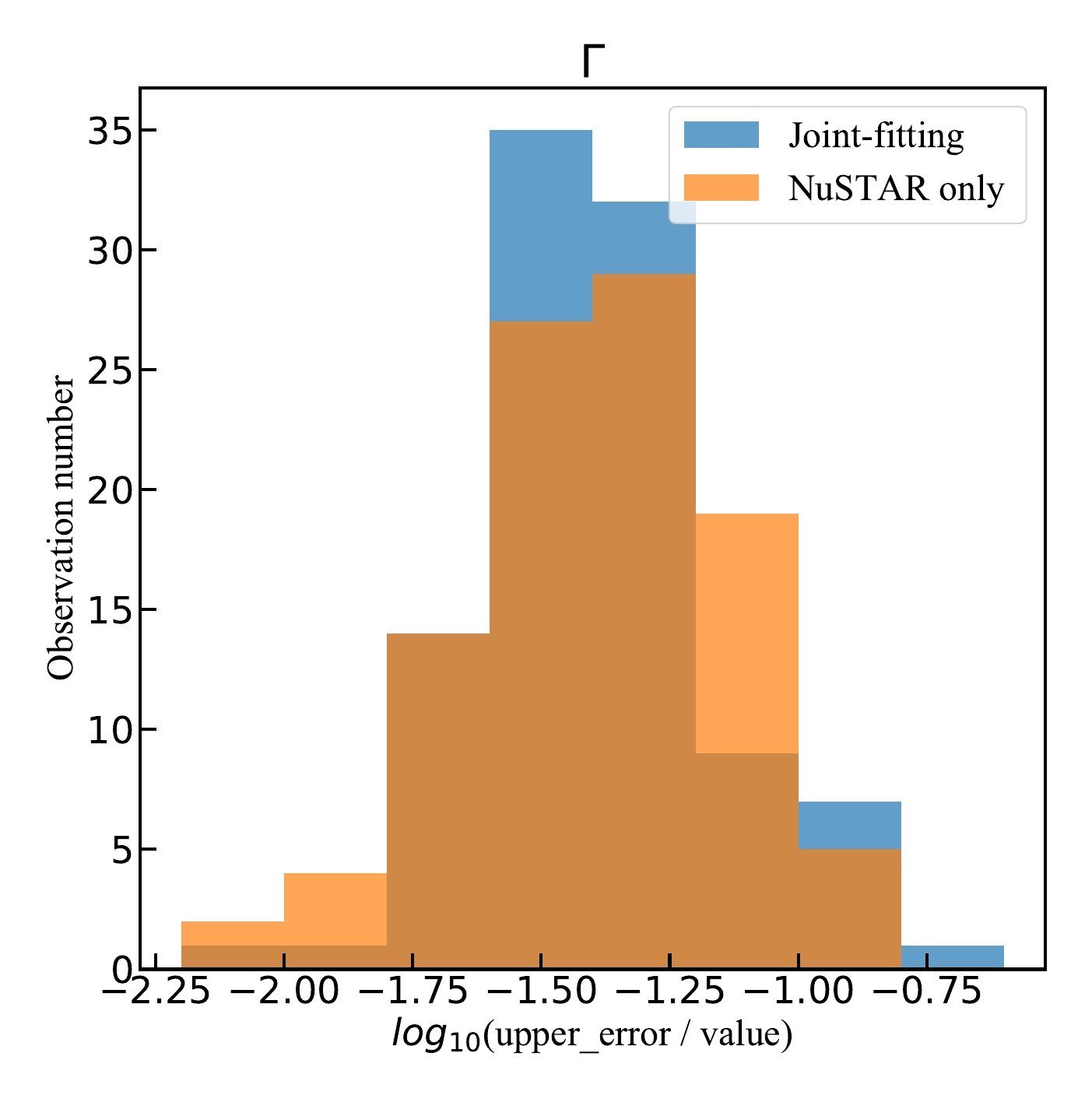} }
\subfloat{\includegraphics[width=0.33\textwidth]{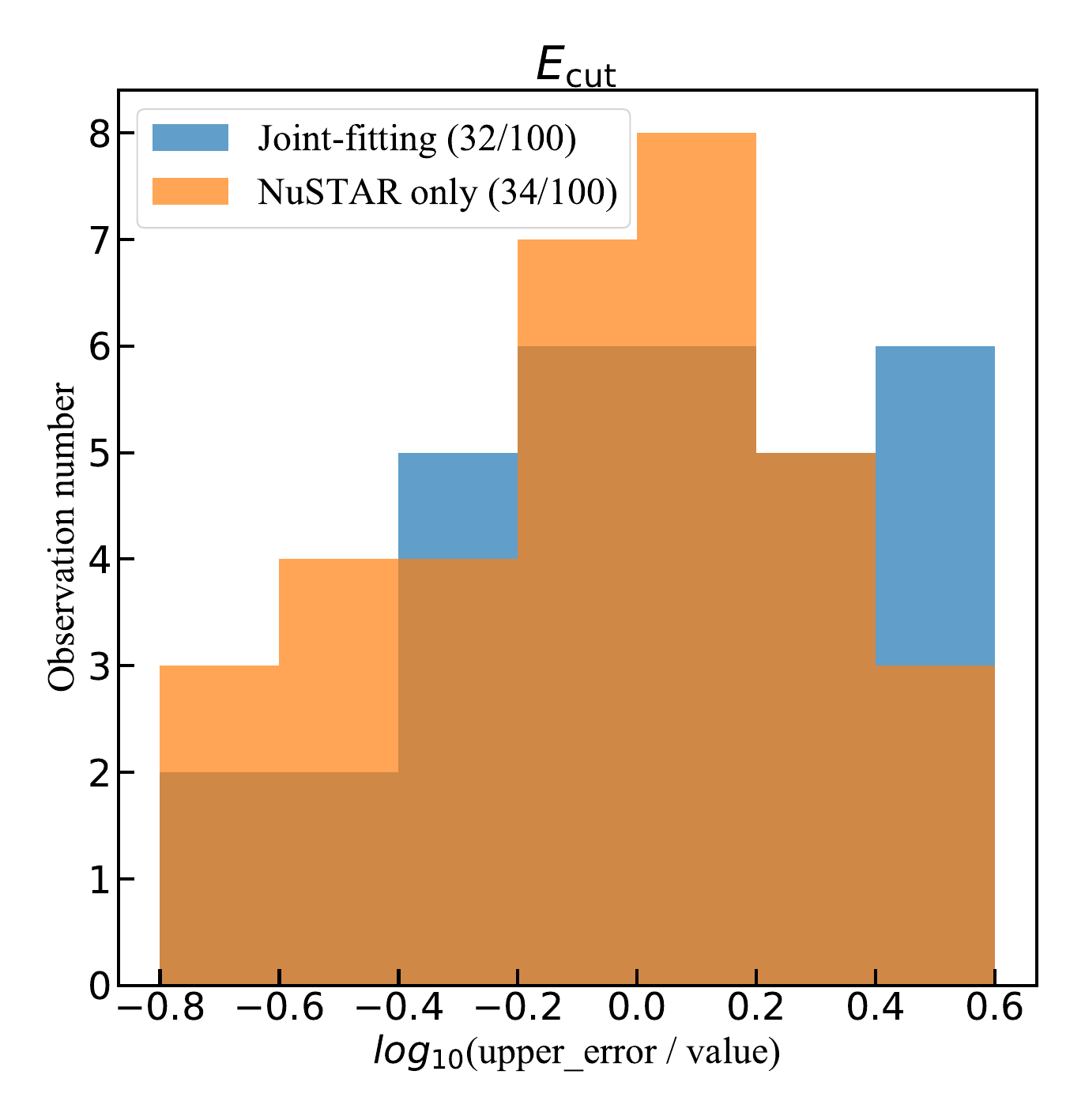} }
\subfloat{\includegraphics[width=0.33\textwidth]{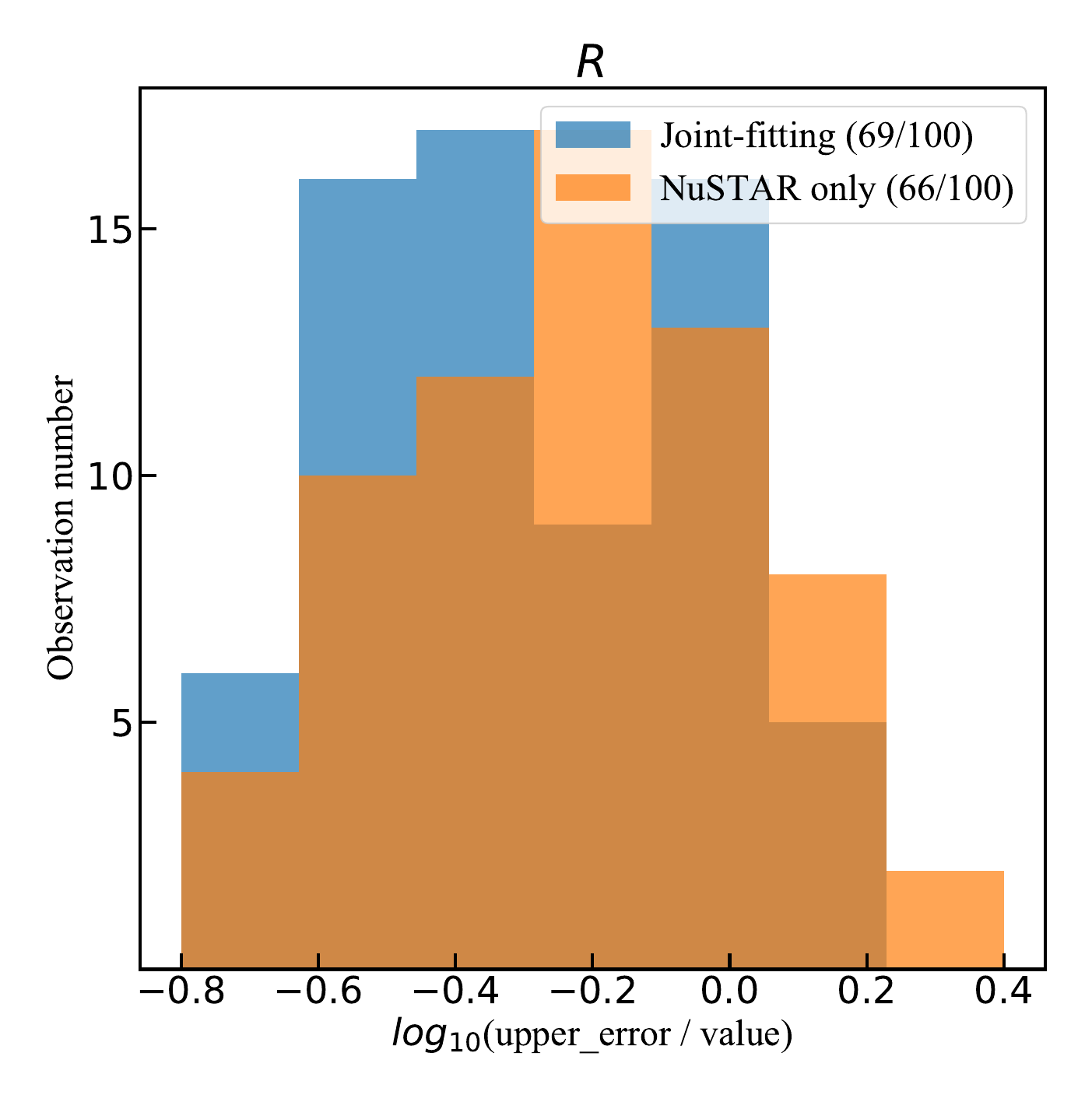} }\\
\subfloat{\includegraphics[width=0.33\textwidth]{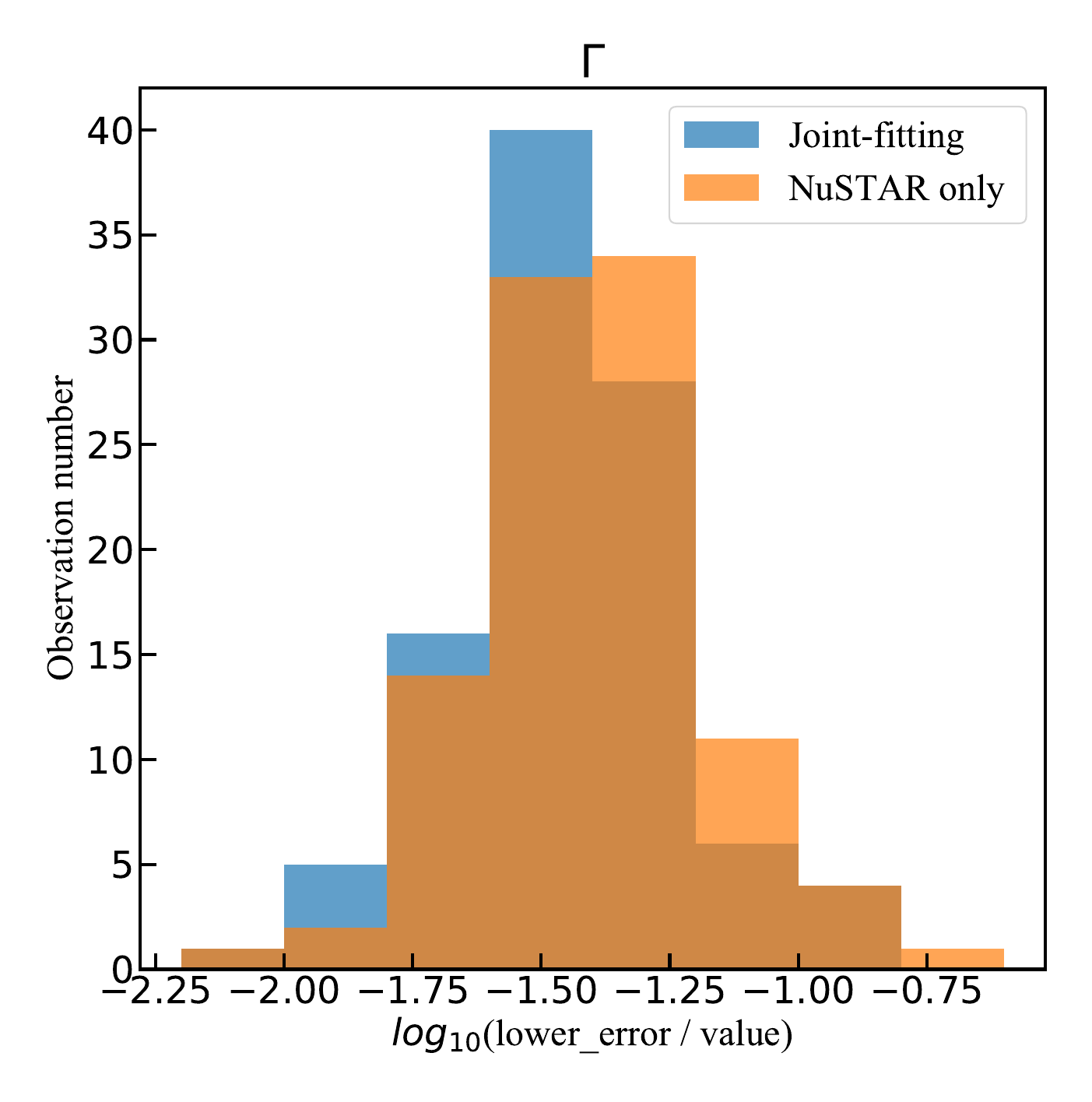} }
\subfloat{\includegraphics[width=0.33\textwidth]{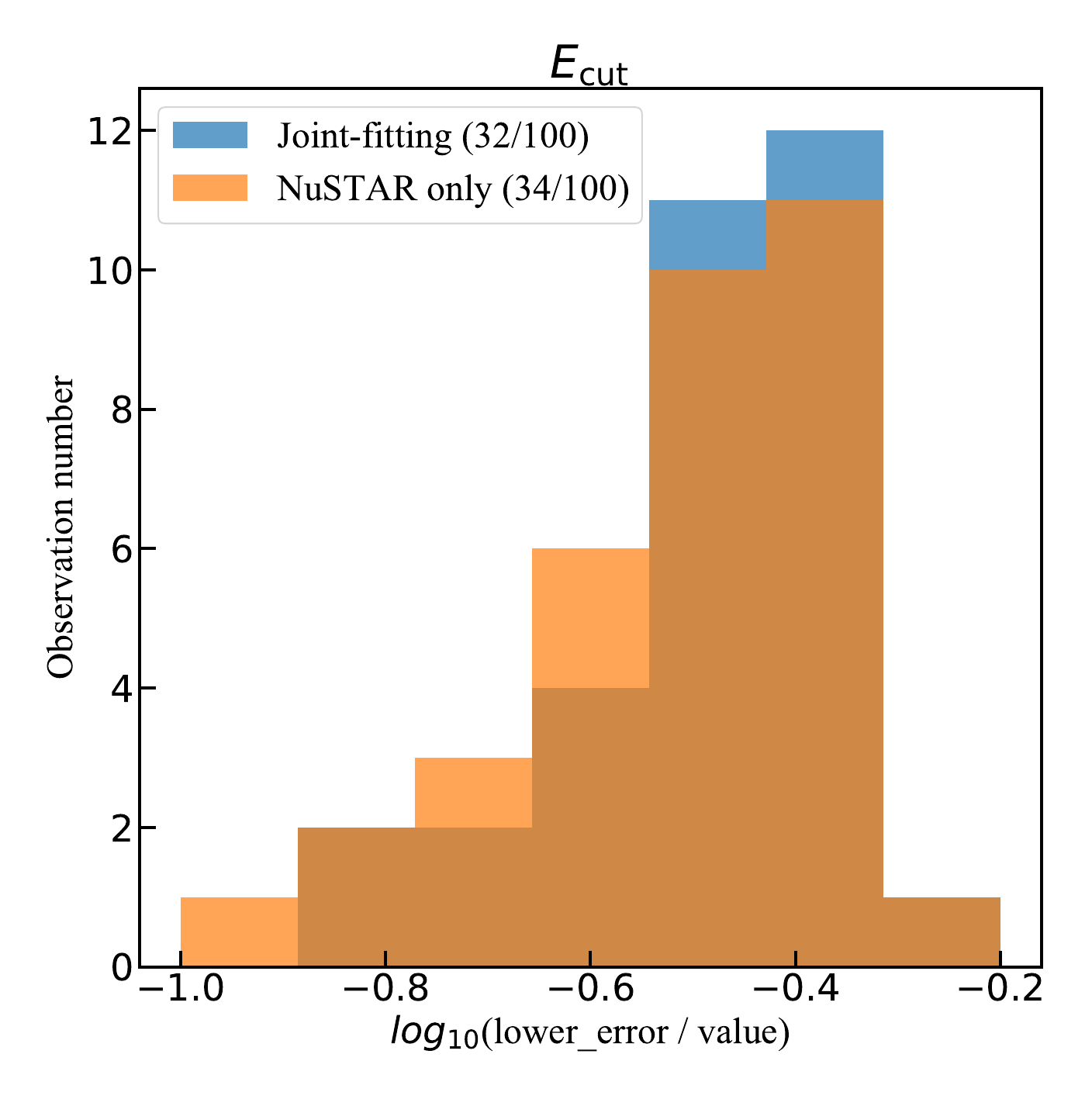} }
\subfloat{\includegraphics[width=0.33\textwidth]{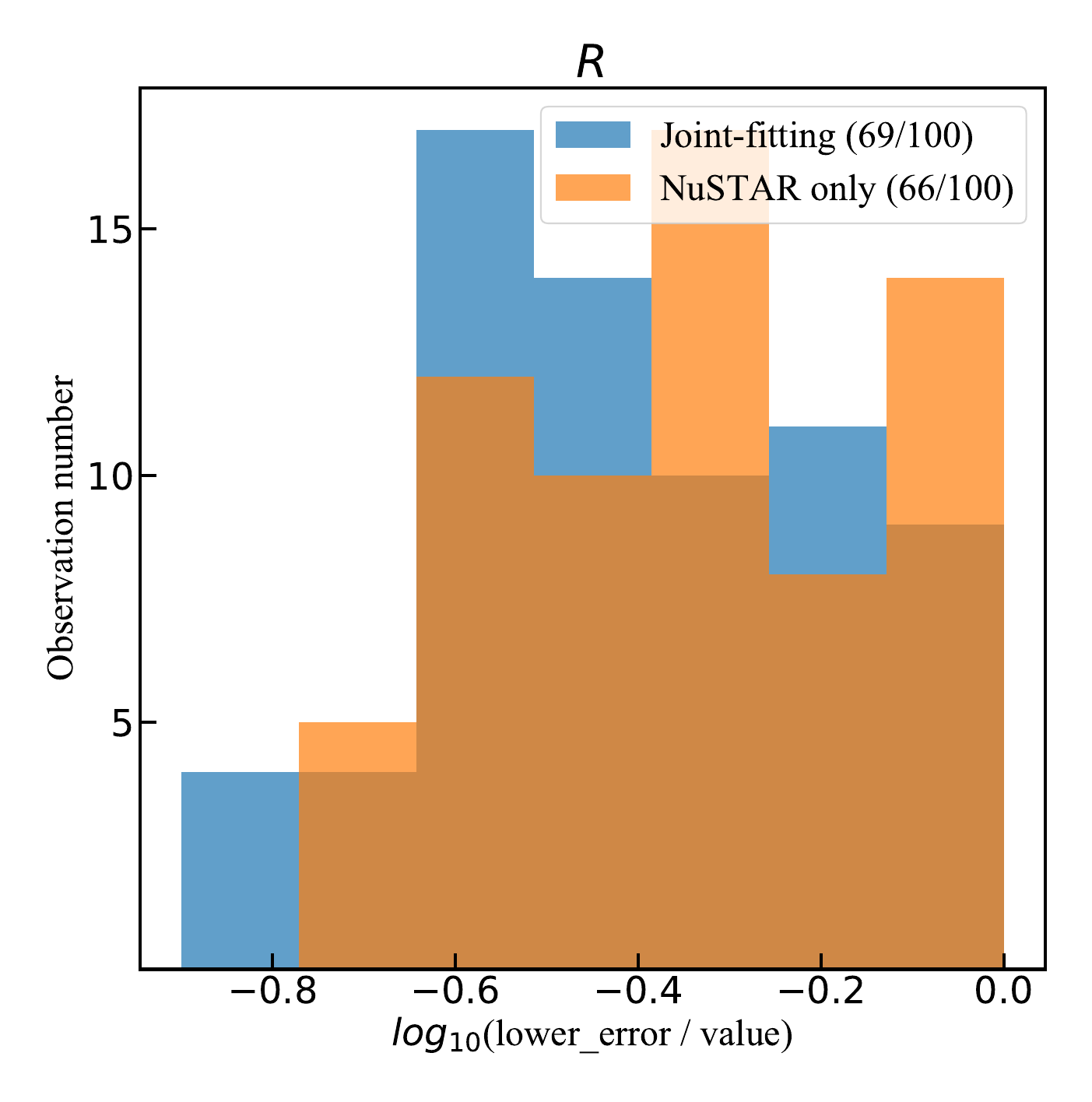} }
\caption{\label{fig:effectiveness} The distribution of the relative errors of $\Gamma$, $R$ and $E_{\rm cut}$, with the detection fraction of $R$ and $E_{\rm cut}$ provided in the legend. Blue boxes show the results of the joint-fitting of \Nu and EPIC-pn, while the orange boxes show the result of fitting the \Nu spectra only (from the whole \Nu exposure without matching EPIC-pn GTI). }
\end{figure*}

\par As shown above, applying the \XMM released ARF correction is highly effective, making the joint-fitting of \Nu and \XMM spectra valid and feasible. However, compared with \citet{Kang_2022} which fits the \Nu spectra alone, it seems that the joint-fitting in this work does not significantly improve the constraints to the spectral parameters. Due to various observational restrictions, the proposed coordinated observations of \Nu and \XMM often can not perfectly overlap (see Figure \ref{fig:Mrk1383}). Moreover, filtering the periods with flaring background for \XMM also loses exposure time (see also Figure \ref{fig:Mrk1383}). Therefore, if one requires a perfect simultaneity between \Nu and \XMM data as this work does, the involvement of \XMM data leads to loss of \Nu exposure time and hence does not always improve the fitting. {As shown in Table \ref{tab:sole} and Figure \ref{fig:explost}, about 50\% of the observations will lose more than 50\% of the \Nu exposure time, if requiring a perfect simultaneity between \Nu and \XMM data. We illustrate an example of Mrk 1383 in Figure \ref{fig:Mrk1383}, of which a 100 ks coordinated \Nu and \XMM observation was proposed. We first note that for a long \Nu observation in science mode, only about 50\% of the duration time is available for exposure due to the Earth occultation. Therefore, a proposed 100 ks \Nu exposure is acutally discontinuously distributed in a $\sim$ 200 ks duration, indicating it can never be perfectly simultaneous with a 100 ks long \XMM exposure. In the example we show in Figure \ref{fig:Mrk1383} for Mrk 1383, the unique distribution of \Nu exposure directly causes a loss of $\sim$ 50\% net exposure time for both \Nu and {\it XMM-Newton} if requiring a perfect simultaneity. Furthermore, filtering the intervals with flaring background for \XMM causes a further loss of the exposure time. Taken together, inclusion of the \XMM data causes a loss of more than 60\% of the \Nu net exposure time. Note for the Small Window mode of EPIC-pn, the livetime fraction is only 71\%. This leads to a shorter net exposure time of \XMM (28 ks) than \Nu one (38 ks) after merging the GTIs, which further reduces the statistical significance of \XMM data. }

\par To conduct a quantified assessment, we extract \Nu spectra in all the available GTI (without a match with EPIC-pn) and fit them with the same model, the result of which is shown in Table \ref{tab:sole}. In Figure \ref{fig:effectiveness}, we show the distribution of the relative errors of $\Gamma$, $R$ and $E_{\rm cut}$, along with the detection fraction of $R$ and $E_{\rm cut}$. In conclusion, for this sample and this model, the involvement of \XMM data does not significantly help. 

\par If we make a concession about the simultaneity, simply including a quasi-simultaneous \XMM exposure will certainly improve the constraint of the spectral parameters, which however could bring biases for variable sources like AGN. Moreover, here we adopt $pexrav$, a simple model without strong physical assumptions, to model the reflection component, while models like $pexmon$ \citep{Nandra_2007}, $xillver$ and $relxill$ \citep{Dauser_2010, Garcia_2014} perform self-consistently fitting on the reflection component and the Fe K$\alpha$ line. EPIC spectra will be statistically more significant on the constraint of \ec and $R$ when fitting with these models, as they provide good constraint for the Fe K$\alpha$ line. However, we stress the coupling between the reflection component and Fe K$\alpha$ line is yet unclear \citep{Chiang_2000, Mantovani_2016, Kang_2020}. Finally, we note the current correction of the effective area is limited to 3--12 keV, indicating the EPIC data below 3 keV are not usable when performing joint-fitting with {\it NuSTAR}. With proper soft band correction and wider dynamic range, EPIC spectral would be statistically more important, though the complicated absorption and soft excess in the soft band might induce other biases.

\section{Summary} \label{sec:summary}
\par In this work we perform joint-fitting of \Nu and \XMM EPIC-pn spectra for a large sample of 104 observation pairs of 44 AGN. Below are our main results.
 
\begin{enumerate}
\item Calibration issue does exist between two missions; EPIC-pn spectra are systematically harder than those of \Nu ($\Delta \Gamma \sim 0.1$), leading to underestimated cutoff energy \ec and reflection component $R$ when performing joint-fitting before correcting the calibration issue. 
\item The empirical correction of the effective area implemented in latest \XMM calibration files (but would not be applied by default) is highly effective and could commendably erase the discrepancy in the derived best-fit $\Gamma$, \ec and $R$.
\item For this sample, requiring a perfect simultaneity between the \Nu and EPIC-pn spectra leads to serious loss of net exposure time of {\it NuSTAR}. Consequently, fitting \Nu spectra jointly with simultaneous EPIC-pn data does not always improve the constraints to the key spectral parameters. 
\item For \XMM EPIC-pn observations in Small Window mode, insufficient filtering of high background flares could bias the spectral fitting results due to the background vignetting effect, which is no longer negligible in case of background flares.
A threshold of 0.05 counts/s to filter background flares 
(see \S \ref{sec:data} for the definition) appears appropriate for EPIC-pn Small Window mode. 
\end{enumerate}

Finally, we note again the correction of the effective area is not applied by default in $arfgen$. Currently, the correction is limited to 3--12 keV and thus not applicable if the soft X-ray band data are included. 
If a similar calibration issue exists below 3 keV, the twisted response curve itself could bias the measurement of the soft excess and ionized absorption, even when fitting the EPIC-pn spectrum alone. We look forward to an updated correction curve in future calibration releases.  
 
 \acknowledgments
 {This research has made use of the {\it NuSTAR} Data Analysis Software (NuSTARDAS) jointly developed by the ASI Science Data Center (ASDC, Italy) and the California Institute of Technology (USA). This work is based on observations obtained with {\it XMM-Newton}, an ESA science mission with instruments and contributions directly funded by ESA Member States and NASA. This work is supported by National Natural Science Foundation of China (grants No. 11890693, 12033006 $\&$ 12192221). The authors gratefully acknowledge the support of Cyrus Chun Ying Tang Foundations.
 }

 \newpage
 
 \startlongtable
 \begin{deluxetable*}{ccccccccccccc}
\tabletypesize{\scriptsize}
\renewcommand\tabcolsep{2.1pt}
\tablecaption{Fitting results of NuSTAR, XMM, and joint spectra (requiring perfect simultaneity between NuSTAR and XMM exposures) \label{tab:obs}}
\tablehead{
\colhead{Source} & \colhead{Nu ID} & \colhead{XMM ID} & \colhead{Mode} & \colhead{$\Gamma^{\rm Nu}$} & \colhead{$\Gamma^{\rm pn}$} & \colhead{$\Gamma^{\rm pn-Cor}$} & \colhead{$E_{\rm cut}^{\rm Nu}$} & \colhead{$E_{\rm cut}^{\rm joint}$} & \colhead{$E_{\rm cut}^{\rm joint-Cor }$} & \colhead{$R^{\rm Nu}$} & \colhead{$R^{\rm joint}$} & \colhead{$R^{\rm joint-Cor }$} \\
\colhead{} & \colhead{} & \colhead{} & \colhead{} & \colhead{} & \colhead{} & \colhead{} & \colhead{keV} & \colhead{keV} & \colhead{keV} & \colhead{} & \colhead{} & \colhead{} 
}  
\startdata
Fairall 9 & 60001130002 & 0741330101 & SW & $1.78_{-0.21}^{+0.16}$ & $1.86_{-0.20}^{+0.22}$ & $1.97_{-0.20}^{+0.24}$ & $ > 38 $ & $ > 37 $ & $ > 38 $ & $ < 1.05 $ & $ < 1.34 $ & $ < 1.67 $ \\
Mrk 359 & 60402021004 & 0830550901 & SW & $1.84_{-0.12}^{+0.14}$ & $1.71_{-0.09}^{+0.09}$ & $1.79_{-0.09}^{+0.09}$ & $ > 95 $ & $ > 89 $ & $ > 99 $ & $0.85_{-0.58}^{+0.97}$ & $0.56_{-0.44}^{+0.72}$ & $0.72_{-0.46}^{+0.79}$ \\
 & 60402021006 & 0830551001 & SW & $1.84_{-0.11}^{+0.04}$ & $1.75_{-0.09}^{+0.09}$ & $1.82_{-0.08}^{+0.09}$ & $ > 109 $ & $ > 112 $ & $ > 125 $ & $0.53_{-0.43}^{+0.72}$ & $0.32_{-0.31}^{+0.56}$ & $0.47_{-0.35}^{+0.59}$ \\
 & 60402021008 & 0830551101 & SW & $1.78_{-0.24}^{+0.27}$ & $1.50_{-0.09}^{+0.10}$ & $1.57_{-0.08}^{+0.10}$ & $ > 65 $ & $ > 52 $ & $ > 49 $ & $ < 1.97 $ & $ < 1.02 $ & $ < 1.19 $ \\
NGC 931 & 60101002002 & 0760530201 & LW & $1.88_{-0.08}^{+0.07}$ & $1.75_{-0.05}^{+0.05}$ & $1.86_{-0.05}^{+0.05}$ & $ > 260 $ & $ > 195 $ & $ > 248 $ & $0.73_{-0.23}^{+0.27}$ & $0.56_{-0.18}^{+0.20}$ & $0.78_{-0.21}^{+0.23}$ \\
 & 60101002004 & 0760530301 & LW & $1.89_{-0.06}^{+0.08}$ & $1.81_{-0.03}^{+0.03}$ & $1.88_{-0.03}^{+0.05}$ & $ > 367 $ & $ > 259 $ & $ > 324 $ & $0.80_{-0.25}^{+0.33}$ & $0.60_{-0.18}^{+0.24}$ & $0.84_{-0.21}^{+0.29}$ \\
Mrk 1044 & 60401005002 & 0824080301 & SW & $2.50_{-0.09}^{+0.10}$ & $2.42_{-0.04}^{+0.04}$ & $2.48_{-0.04}^{+0.06}$ & $ > 126 $ & $ > 123 $ & $ > 146 $ & $1.75_{-0.50}^{+0.61}$ & $1.31_{-0.37}^{+0.43}$ & $1.52_{-0.39}^{+0.46}$ \\
 & 60401005002 & 0824080501 & SW & $2.32_{-0.08}^{+0.08}$ & $2.27_{-0.06}^{+0.05}$ & $2.34_{-0.05}^{+0.04}$ & $ > 129 $ & $ > 129 $ & $ > 144 $ & $0.83_{-0.41}^{+0.62}$ & $1.01_{-0.49}^{+0.46}$ & $1.06_{-0.28}^{+0.63}$ \\
3C 109 & 60301011002 & 0795600101 & LW & $1.68_{-0.17}^{+0.12}$ & $1.69_{-0.16}^{+0.18}$ & $1.79_{-0.18}^{+0.18}$ & $ > 67 $ & $ > 74 $ & $ > 85 $ & $ < 0.32 $ & $ < 0.55 $ & $ < 0.40 $ \\
NGC 1566 & 80301601002 & 0800840201 & SW & $1.84_{-0.05}^{+0.03}$ & $1.77_{-0.02}^{+0.02}$ & $1.85_{-0.02}^{+0.03}$ & $ > 610 $ & $ > 351 $ & $ > 475 $ & $0.74_{-0.15}^{+0.16}$ & $0.60_{-0.11}^{+0.12}$ & $0.79_{-0.13}^{+0.15}$ \\
 & 80401601002 & 0820530401 & SW & $1.77_{-0.09}^{+0.08}$ & $1.60_{-0.03}^{+0.04}$ & $1.69_{-0.04}^{+0.06}$ & $ > 241 $ & $ > 158 $ & $ > 182 $ & $0.67_{-0.25}^{+0.29}$ & $0.49_{-0.20}^{+0.23}$ & $0.64_{-0.22}^{+0.24}$ \\
 & 80502606002 & 0840800401 & SW & $1.78_{-0.12}^{+0.04}$ & $1.62_{-0.04}^{+0.04}$ & $1.69_{-0.04}^{+0.04}$ & $ > 182 $ & $ > 106 $ & $ > 117 $ & $0.42_{-0.28}^{+0.36}$ & $0.25_{-0.23}^{+0.27}$ & $0.36_{-0.25}^{+0.30}$ \\
1H 0419-577 & 60402006002 & 0820360101 & SW & $1.65_{-0.17}^{+0.14}$ & $1.55_{-0.04}^{+0.05}$ & $1.62_{-0.04}^{+0.05}$ & $63_{-24}^{+46}$ & $52_{-13}^{+31}$ & $56_{-18}^{+35}$ & $ < 0.55 $ & $ < 0.35 $ & $ < 0.53 $ \\
 & 60402006004 & 0820360201 & SW & $1.55_{-0.13}^{+0.19}$ & $1.50_{-0.05}^{+0.05}$ & $1.57_{-0.05}^{+0.05}$ & $54_{-19}^{+58}$ & $49_{-14}^{+21}$ & $50_{-14}^{+35}$ & $ < 0.52 $ & $ < 0.40 $ & $ < 0.60 $ \\
Ark 120 & 60001044002 & 0693781501 & SW & $1.83_{-0.04}^{+0.08}$ & $1.79_{-0.06}^{+0.06}$ & $1.90_{-0.06}^{+0.06}$ & $ > 612 $ & $ > 714 $ & $ > 881 $ & $0.49_{-0.17}^{+0.27}$ & $0.54_{-0.19}^{+0.22}$ & $0.77_{-0.22}^{+0.26}$ \\
 & 60001044004 & 0721600401 & SW & $1.98_{-0.03}^{+0.05}$ & $1.90_{-0.02}^{+0.02}$ & $1.97_{-0.02}^{+0.02}$ & $ > 346 $ & $ > 266 $ & $ > 335 $ & $0.69_{-0.18}^{+0.18}$ & $0.50_{-0.07}^{+0.15}$ & $0.69_{-0.15}^{+0.17}$ \\
ESO 362-18 & 60201046002 & 0790810101 & SW & $1.48_{-0.12}^{+0.14}$ & $1.33_{-0.05}^{+0.04}$ & $1.40_{-0.05}^{+0.06}$ & $119_{-45}^{+149}$ & $95_{-28}^{+67}$ & $101_{-31}^{+75}$ & $0.56_{-0.31}^{+0.39}$ & $0.42_{-0.24}^{+0.29}$ & $0.57_{-0.27}^{+0.32}$ \\
2MASX J05210136-2521450 & 60201022002 & 0790580101 & FF & $2.11_{-0.17}^{+0.17}$ & $2.12_{-0.14}^{+0.14}$ & $2.24_{-0.14}^{+0.14}$ & $ > 114 $ & $ > 103 $ & $ > 115 $ & $ < 0.87 $ & $ < 0.94 $ & $0.57_{-0.47}^{+0.66}$ \\
Mrk 79 & 60601010004 & 0870880101 & SW & $1.77_{-0.08}^{+0.16}$ & $1.77_{-0.05}^{+0.05}$ & $1.86_{-0.06}^{+0.11}$ & $ > 64 $ & $121_{-54}^{+575}$ & $ > 75 $ & $0.62_{-0.42}^{+0.61}$ & $0.61_{-0.40}^{+0.50}$ & $0.80_{-0.43}^{+0.55}$ \\
Mrk 110 & 60502022002 & 0852590101 & SW & $1.87_{-0.13}^{+0.13}$ & $1.70_{-0.05}^{+0.05}$ & $1.78_{-0.06}^{+0.11}$ & $ > 114 $ & $ > 77 $ & $ > 94 $ & $ < 0.39 $ & $ < 0.19 $ & $ < 0.25 $ \\
 & 60502022004 & 0852590201 & SW & $1.87_{-0.11}^{+0.11}$ & $1.74_{-0.04}^{+0.08}$ & $1.85_{-0.07}^{+0.08}$ & $ > 111 $ & $ > 86 $ & $ > 93 $ & $0.38_{-0.30}^{+0.37}$ & $ < 0.52 $ & $0.39_{-0.27}^{+0.32}$ \\
NGC 2992 & 90501623002 & 0840920301 & SW & $1.67_{-0.04}^{+0.04}$ & $1.56_{-0.03}^{+0.03}$ & $1.67_{-0.03}^{+0.03}$ & $353_{-134}^{+545}$ & $302_{-99}^{+278}$ & $377_{-138}^{+515}$ & $ < 0.16 $ & $ < 0.09 $ & $0.13_{-0.07}^{+0.08}$ \\
MCG -05-23-016 & 60701014002 & 0890670101 & SW & $1.83_{-0.05}^{+0.05}$ & $1.80_{-0.03}^{+0.03}$ & $1.91_{-0.03}^{+0.03}$ & $110_{-19}^{+29}$ & $109_{-17}^{+24}$ & $124_{-21}^{+32}$ & $0.77_{-0.15}^{+0.17}$ & $0.72_{-0.13}^{+0.14}$ & $0.96_{-0.15}^{+0.17}$ \\
NGC 3227 & 60202002002 & 0782520201 & SW & $1.80_{-0.10}^{+0.10}$ & $1.61_{-0.06}^{+0.06}$ & $1.72_{-0.06}^{+0.06}$ & $189_{-73}^{+260}$ & $147_{-44}^{+104}$ & $171_{-56}^{+147}$ & $1.20_{-0.33}^{+0.40}$ & $0.84_{-0.23}^{+0.27}$ & $1.03_{-0.26}^{+0.30}$ \\
 & 60202002004 & 0782520301 & SW & $1.61_{-0.11}^{+0.13}$ & $1.52_{-0.07}^{+0.07}$ & $1.64_{-0.07}^{+0.07}$ & $98_{-29}^{+64}$ & $104_{-28}^{+57}$ & $115_{-33}^{+69}$ & $0.88_{-0.33}^{+0.41}$ & $0.83_{-0.27}^{+0.33}$ & $1.02_{-0.30}^{+0.36}$ \\
 & 60202002006 & 0782520401 & SW & $1.82_{-0.08}^{+0.08}$ & $1.62_{-0.05}^{+0.06}$ & $1.73_{-0.06}^{+0.06}$ & $ > 166 $ & $186_{-65}^{+192}$ & $219_{-83}^{+290}$ & $0.94_{-0.27}^{+0.32}$ & $0.66_{-0.20}^{+0.23}$ & $0.83_{-0.22}^{+0.26}$ \\
 & 60202002008 & 0782520501 & SW & $1.86_{-0.07}^{+0.05}$ & $1.69_{-0.04}^{+0.05}$ & $1.80_{-0.05}^{+0.05}$ & $ > 510 $ & $ > 262 $ & $ > 330 $ & $0.89_{-0.23}^{+0.25}$ & $0.63_{-0.17}^{+0.19}$ & $0.79_{-0.19}^{+0.21}$ \\
 & 60202002010 & 0782520601 & SW & $1.81_{-0.06}^{+0.07}$ & $1.70_{-0.03}^{+0.04}$ & $1.81_{-0.04}^{+0.04}$ & $244_{-89}^{+312}$ & $207_{-64}^{+156}$ & $255_{-88}^{+256}$ & $0.94_{-0.23}^{+0.26}$ & $0.79_{-0.17}^{+0.20}$ & $0.99_{-0.20}^{+0.22}$ \\
 & 60202002012 & 0782520701 & SW & $1.92_{-0.08}^{+0.07}$ & $1.77_{-0.05}^{+0.05}$ & $1.88_{-0.05}^{+0.05}$ & $ > 322 $ & $ > 234 $ & $ > 295 $ & $1.06_{-0.28}^{+0.33}$ & $0.79_{-0.21}^{+0.23}$ & $0.98_{-0.23}^{+0.26}$ \\
 & 80502609002 & 0844341301 & SW & $1.60_{-0.10}^{+0.10}$ & $1.68_{-0.07}^{+0.07}$ & $1.79_{-0.07}^{+0.07}$ & $167_{-69}^{+319}$ & $ > 132 $ & $ > 149 $ & $0.33_{-0.21}^{+0.25}$ & $0.48_{-0.21}^{+0.24}$ & $0.65_{-0.23}^{+0.27}$ \\
 & 80502609004 & 0844341401 & SW & $1.61_{-0.23}^{+0.23}$ & $1.35_{-0.14}^{+0.15}$ & $1.46_{-0.14}^{+0.15}$ & $ > 68 $ & $84_{-34}^{+142}$ & $90_{-37}^{+161}$ & $0.68_{-0.51}^{+0.77}$ & $0.45_{-0.39}^{+0.54}$ & $0.61_{-0.43}^{+0.61}$ \\
2MASSi J1031543-141651 & 60701046002 & 0890410101 & SW & $1.79_{-0.07}^{+0.13}$ & $1.75_{-0.04}^{+0.04}$ & $1.82_{-0.04}^{+0.07}$ & $153_{-62}^{+412}$ & $143_{-54}^{+229}$ & $169_{-70}^{+340}$ & $0.70_{-0.33}^{+0.41}$ & $0.55_{-0.26}^{+0.31}$ & $0.74_{-0.29}^{+0.35}$ \\
NGC 3516 & 60160001002 & 0854591101 & SW & $1.70_{-0.15}^{+0.19}$ & $1.69_{-0.11}^{+0.11}$ & $1.80_{-0.11}^{+0.11}$ & $ > 90 $ & $ > 106 $ & $ > 117 $ & $0.77_{-0.50}^{+0.69}$ & $0.86_{-0.44}^{+0.58}$ & $1.09_{-0.50}^{+0.66}$ \\
HE 1136-2304 & 80002031002 & 0741260101 & SW & $1.78_{-0.16}^{+0.16}$ & $1.65_{-0.11}^{+0.11}$ & $1.76_{-0.11}^{+0.11}$ & $ > 85 $ & $ > 78 $ & $ > 84 $ & $ < 0.79 $ & $ < 0.54 $ & $ < 0.72 $ \\
 & 80002031003 & 0741260101 & SW & $1.60_{-0.09}^{+0.09}$ & $1.51_{-0.06}^{+0.06}$ & $1.58_{-0.06}^{+0.06}$ & $110_{-55}^{+376}$ & $85_{-35}^{+91}$ & $100_{-47}^{+269}$ & $ < 0.50 $ & $ < 0.34 $ & $ < 0.48 $ \\
KUG 1141+371 & 90601618002 & 0871190101 & SW & $1.79_{-0.23}^{+0.33}$ & $1.70_{-0.08}^{+0.08}$ & $1.77_{-0.08}^{+0.13}$ & $40_{-17}^{+102}$ & $39_{-14}^{+67}$ & $43_{-16}^{+71}$ & $0.94_{-0.88}^{+1.50}$ & $ < 1.39 $ & $ < 1.59 $ \\
2MASX J11454045-1827149 & 60302002002 & 0795580101 & SW & $1.74_{-0.14}^{+0.15}$ & $1.63_{-0.05}^{+0.10}$ & $1.74_{-0.08}^{+0.10}$ & $64_{-22}^{+62}$ & $58_{-18}^{+44}$ & $61_{-19}^{+49}$ & $0.51_{-0.39}^{+0.51}$ & $0.32_{-0.31}^{+0.38}$ & $0.46_{-0.34}^{+0.42}$ \\
 & 60302002004 & 0795580201 & SW & $1.67_{-0.09}^{+0.18}$ & $1.54_{-0.06}^{+0.06}$ & $1.61_{-0.06}^{+0.06}$ & $ > 54 $ & $95_{-42}^{+253}$ & $111_{-53}^{+367}$ & $ < 0.84 $ & $ < 0.34 $ & $ < 0.45 $ \\
 & 60302002006 & 0795580301 & SW & $1.75_{-0.08}^{+0.16}$ & $1.70_{-0.04}^{+0.04}$ & $1.77_{-0.04}^{+0.04}$ & $126_{-53}^{+593}$ & $121_{-48}^{+239}$ & $138_{-58}^{+357}$ & $ < 0.64 $ & $ < 0.41 $ & $0.26_{-0.25}^{+0.30}$ \\
 & 60302002008 & 0795580401 & SW & $1.89_{-0.15}^{+0.14}$ & $1.72_{-0.04}^{+0.04}$ & $1.82_{-0.07}^{+0.10}$ & $ > 91 $ & $ > 69 $ & $ > 74 $ & $0.43_{-0.39}^{+0.50}$ & $ < 0.64 $ & $0.41_{-0.34}^{+0.42}$ \\
 & 60302002010 & 0795580501 & SW & $1.90_{-0.19}^{+0.15}$ & $1.73_{-0.05}^{+0.05}$ & $1.79_{-0.03}^{+0.10}$ & $ > 78 $ & $ > 67 $ & $ > 69 $ & $0.50_{-0.49}^{+0.56}$ & $ < 0.54 $ & $ < 0.75 $ \\
NGC 4051 & 60401009002 & 0830430201 & SW & $1.97_{-0.09}^{+0.09}$ & $1.86_{-0.04}^{+0.05}$ & $1.94_{-0.04}^{+0.07}$ & $ > 791 $ & $ > 446 $ & $ > 588 $ & $1.67_{-0.48}^{+0.60}$ & $1.36_{-0.39}^{+0.48}$ & $1.61_{-0.41}^{+0.53}$ \\
NGC 4593 & 60001149002 & 0740920201 & SW & $1.87_{-0.13}^{+0.18}$ & $1.75_{-0.06}^{+0.10}$ & $1.82_{-0.06}^{+0.13}$ & $ > 132 $ & $ > 105 $ & $ > 115 $ & $0.88_{-0.54}^{+0.92}$ & $0.64_{-0.44}^{+0.67}$ & $0.83_{-0.47}^{+0.75}$ \\
 & 60001149004 & 0740920301 & SW & $1.73_{-0.20}^{+0.25}$ & $1.53_{-0.05}^{+0.23}$ & $1.64_{-0.14}^{+0.23}$ & $ > 142 $ & $ > 119 $ & $ > 129 $ & $ < 1.42 $ & $ < 1.07 $ & $ < 1.30 $ \\
 & 60001149006 & 0740920401 & SW & $1.79_{-0.11}^{+0.12}$ & $1.62_{-0.06}^{+0.06}$ & $1.69_{-0.06}^{+0.06}$ & $ > 204 $ & $ > 98 $ & $ > 114 $ & $1.18_{-0.56}^{+0.89}$ & $0.79_{-0.47}^{+0.64}$ & $1.00_{-0.51}^{+0.73}$ \\
 & 60001149008 & 0740920501 & SW & $1.85_{-0.08}^{+0.09}$ & $1.71_{-0.06}^{+0.07}$ & $1.78_{-0.06}^{+0.07}$ & $ > 612 $ & $ > 489 $ & $ > 545 $ & $0.67_{-0.35}^{+0.49}$ & $0.37_{-0.25}^{+0.32}$ & $0.51_{-0.27}^{+0.35}$ \\
 & 60001149010 & 0740920601 & SW & $1.83_{-0.10}^{+0.12}$ & $1.71_{-0.06}^{+0.05}$ & $1.78_{-0.05}^{+0.07}$ & $ > 165 $ & $ > 127 $ & $ > 147 $ & $0.40_{-0.31}^{+0.46}$ & $ < 0.59 $ & $0.36_{-0.26}^{+0.39}$ \\
MCG -06-30-015 & 60001047003 & 0693781201 & SW & $2.33_{-0.06}^{+0.05}$ & $2.26_{-0.04}^{+0.04}$ & $2.37_{-0.04}^{+0.04}$ & $ > 576 $ & $ > 356 $ & $ > 635 $ & $1.71_{-0.30}^{+0.35}$ & $1.43_{-0.23}^{+0.26}$ & $1.77_{-0.27}^{+0.27}$ \\
 & 60001047003 & 0693781301 & SW & $2.24_{-0.05}^{+0.05}$ & $2.10_{-0.04}^{+0.04}$ & $2.21_{-0.04}^{+0.04}$ & $ > 693 $ & $ > 318 $ & $ > 503 $ & $1.75_{-0.31}^{+0.35}$ & $1.41_{-0.23}^{+0.26}$ & $1.80_{-0.26}^{+0.32}$ \\
 & 60001047005 & 0693781401 & SW & $2.12_{-0.12}^{+0.11}$ & $2.07_{-0.09}^{+0.10}$ & $2.18_{-0.09}^{+0.10}$ & $ > 281 $ & $ > 326 $ & $ > 425 $ & $1.87_{-0.66}^{+0.88}$ & $1.74_{-0.53}^{+0.69}$ & $2.06_{-0.60}^{+0.78}$ \\
IC 4329A & 60702050002 & 0862090101 & SW & $1.82_{-0.11}^{+0.10}$ & $1.75_{-0.07}^{+0.07}$ & $1.86_{-0.07}^{+0.07}$ & $ > 138 $ & $ > 142 $ & $ > 164 $ & $0.58_{-0.27}^{+0.33}$ & $0.51_{-0.23}^{+0.26}$ & $0.68_{-0.25}^{+0.30}$ \\
 & 60702050004 & 0862090301 & SW & $1.74_{-0.12}^{+0.11}$ & $1.67_{-0.08}^{+0.08}$ & $1.78_{-0.08}^{+0.08}$ & $ > 227 $ & $ > 204 $ & $ > 237 $ & $ < 0.54 $ & $ < 0.39 $ & $0.28_{-0.20}^{+0.27}$ \\
 & 60702050006 & 0862090501 & SW & $1.79_{-0.07}^{+0.07}$ & $1.74_{-0.06}^{+0.06}$ & $1.85_{-0.06}^{+0.06}$ & $ > 450 $ & $ > 398 $ & $ > 530 $ & $0.26_{-0.17}^{+0.20}$ & $0.21_{-0.14}^{+0.16}$ & $0.36_{-0.16}^{+0.18}$ \\
 & 60702050010 & 0862090901 & SW & $1.79_{-0.28}^{+0.34}$ & $1.58_{-0.10}^{+0.17}$ & $1.66_{-0.10}^{+0.20}$ & $ > 24 $ & $ > 24 $ & $ > 25 $ & $ < 2.46 $ & $ < 1.33 $ & $ < 1.60 $ \\
NGC 5548 & 60002044003 & 0720110601 & SW & $1.77_{-0.08}^{+0.08}$ & $1.65_{-0.05}^{+0.06}$ & $1.76_{-0.05}^{+0.06}$ & $ > 186 $ & $ > 177 $ & $ > 214 $ & $0.64_{-0.22}^{+0.26}$ & $0.45_{-0.17}^{+0.19}$ & $0.61_{-0.19}^{+0.21}$ \\
 & 60002044005 & 0720111001 & SW & $1.52_{-0.09}^{+0.09}$ & $1.53_{-0.06}^{+0.06}$ & $1.64_{-0.06}^{+0.06}$ & $94_{-23}^{+39}$ & $102_{-23}^{+42}$ & $113_{-28}^{+51}$ & $0.46_{-0.20}^{+0.22}$ & $0.52_{-0.18}^{+0.21}$ & $0.69_{-0.20}^{+0.23}$ \\
 & 60002044008 & 0720111501 & SW & $1.50_{-0.11}^{+0.10}$ & $1.45_{-0.07}^{+0.07}$ & $1.56_{-0.07}^{+0.07}$ & $107_{-31}^{+68}$ & $108_{-28}^{+55}$ & $119_{-32}^{+67}$ & $0.38_{-0.21}^{+0.24}$ & $0.34_{-0.17}^{+0.20}$ & $0.46_{-0.19}^{+0.21}$ \\
ESO 511-G030 & 60502035002 & 0852010101 & SW & $1.72_{-0.26}^{+0.33}$ & $1.56_{-0.10}^{+0.17}$ & $1.60_{-0.10}^{+0.18}$ & $ > 40 $ & $ > 42 $ & $ > 44 $ & $ < 1.51 $ & $ < 0.70 $ & $ < 0.80 $ \\
 & 60502035004 & 0852010201 & SW & $1.73_{-0.22}^{+0.24}$ & $1.51_{-0.10}^{+0.19}$ & $1.58_{-0.09}^{+0.23}$ & $ > 82 $ & $ > 54 $ & $ > 60 $ & $ < 0.37 $ & $ < 0.45 $ & $ < 0.52 $ \\
 & 60502035006 & 0852010301 & SW & $1.45_{-0.17}^{+0.15}$ & $1.47_{-0.10}^{+0.11}$ & $1.54_{-0.09}^{+0.10}$ & $ > 30 $ & $52_{-21}^{+82}$ & $66_{-30}^{+199}$ & $ < 0.76 $ & $ < 0.50 $ & $ < 0.58 $ \\
 & 60502035008 & 0852010401 & SW & $1.57_{-0.16}^{+0.16}$ & $1.54_{-0.10}^{+0.10}$ & $1.61_{-0.10}^{+0.10}$ & $ > 35 $ & $ > 34 $ & $ > 36 $ & $ < 1.23 $ & $ < 1.19 $ & $ < 1.43 $ \\
Mrk 1383 & 60501049002 & 0852210101 & SW & $1.77_{-0.10}^{+0.10}$ & $1.77_{-0.06}^{+0.06}$ & $1.84_{-0.06}^{+0.06}$ & $ > 84 $ & $ > 80 $ & $ > 89 $ & $0.55_{-0.42}^{+0.63}$ & $0.57_{-0.42}^{+0.56}$ & $0.78_{-0.46}^{+0.63}$ \\
Mrk 817 & 60702008002 & 0882340601 & LW & $2.16_{-0.08}^{+0.09}$ & $2.15_{-0.06}^{+0.07}$ & $2.26_{-0.06}^{+0.08}$ & $ > 988 $ & $ > 1112 $ & $ > 1310 $ & $1.57_{-0.44}^{+0.54}$ & $1.36_{-0.33}^{+0.38}$ & $1.64_{-0.36}^{+0.42}$ \\
 & 60702008004 & 0882340701 & LW & $1.70_{-0.16}^{+0.16}$ & $1.77_{-0.10}^{+0.10}$ & $1.88_{-0.10}^{+0.10}$ & $114_{-49}^{+244}$ & $162_{-77}^{+704}$ & $ > 95 $ & $0.95_{-0.45}^{+0.60}$ & $0.92_{-0.38}^{+0.47}$ & $1.16_{-0.42}^{+0.53}$ \\
 & 60702008006 & 0882340801 & LW & $1.89_{-0.18}^{+0.12}$ & $1.99_{-0.11}^{+0.11}$ & $2.10_{-0.11}^{+0.11}$ & $ > 168 $ & $ > 395 $ & $ > 497 $ & $1.65_{-0.42}^{+0.91}$ & $1.74_{-0.56}^{+0.73}$ & $2.05_{-0.62}^{+0.81}$ \\
Mrk 841 & 60101023002 & 0763790501 & FF & $1.88_{-0.13}^{+0.28}$ & $1.94_{-0.13}^{+0.15}$ & $2.06_{-0.14}^{+0.15}$ & $ > 62 $ & $ > 73 $ & $ > 84 $ & $0.99_{-0.70}^{+1.28}$ & $1.02_{-0.65}^{+0.93}$ & $1.27_{-0.71}^{+1.04}$ \\
 & 60702007002 & 0882130301 & SW & $1.85_{-0.11}^{+0.14}$ & $1.80_{-0.04}^{+0.07}$ & $1.91_{-0.07}^{+0.07}$ & $ > 95 $ & $145_{-57}^{+227}$ & $168_{-71}^{+341}$ & $0.95_{-0.41}^{+0.51}$ & $0.90_{-0.33}^{+0.40}$ & $1.20_{-0.38}^{+0.48}$ \\
 & 60702007002 & 0882130401 & SW & $1.90_{-0.09}^{+0.10}$ & $1.82_{-0.04}^{+0.05}$ & $1.90_{-0.04}^{+0.08}$ & $ > 395 $ & $ > 295 $ & $ > 354 $ & $0.66_{-0.31}^{+0.39}$ & $0.50_{-0.23}^{+0.29}$ & $0.67_{-0.24}^{+0.32}$ \\
 & 80701616002 & 0890640201 & SW & $1.87_{-0.13}^{+0.18}$ & $1.75_{-0.04}^{+0.04}$ & $1.84_{-0.06}^{+0.09}$ & $107_{-46}^{+264}$ & $91_{-34}^{+121}$ & $99_{-38}^{+147}$ & $1.19_{-0.57}^{+0.79}$ & $0.84_{-0.42}^{+0.53}$ & $1.10_{-0.47}^{+0.61}$ \\
3C 382 & 60202015002 & 0790600101 & SW & $1.76_{-0.08}^{+0.05}$ & $1.72_{-0.07}^{+0.10}$ & $1.83_{-0.10}^{+0.10}$ & $ > 140 $ & $ > 118 $ & $ > 179 $ & $ < 0.25 $ & $ < 0.24 $ & $ < 0.38 $ \\
 & 60202015004 & 0790600201 & SW & $1.70_{-0.09}^{+0.12}$ & $1.61_{-0.07}^{+0.08}$ & $1.72_{-0.08}^{+0.08}$ & $76_{-23}^{+62}$ & $73_{-21}^{+50}$ & $79_{-24}^{+58}$ & $ < 0.58 $ & $ < 0.43 $ & $0.30_{-0.24}^{+0.29}$ \\
 & 60202015006 & 0790600301 & SW & $1.74_{-0.06}^{+0.07}$ & $1.59_{-0.05}^{+0.08}$ & $1.68_{-0.07}^{+0.07}$ & $ > 120 $ & $144_{-63}^{+373}$ & $ > 98 $ & $ < 0.27 $ & $ < 0.16 $ & $ < 0.18 $ \\
 & 60202015008 & 0790600401 & SW & $1.76_{-0.14}^{+0.15}$ & $1.66_{-0.04}^{+0.05}$ & $1.73_{-0.04}^{+0.04}$ & $ > 86 $ & $133_{-47}^{+152}$ & $190_{-93}^{+541}$ & $ < 0.37 $ & $ < 0.15 $ & $ < 0.27 $ \\
 & 60202015010 & 0790600501 & SW & $1.77_{-0.06}^{+0.05}$ & $1.60_{-0.04}^{+0.04}$ & $1.67_{-0.04}^{+0.04}$ & $ > 138 $ & $140_{-57}^{+278}$ & $ > 97 $ & $ < 0.26 $ & $ < 0.11 $ & $ < 0.14 $ \\
Fairall 49 & 60301028002 & 0795690101 & SW & $2.49_{-0.13}^{+0.12}$ & $2.33_{-0.08}^{+0.08}$ & $2.44_{-0.08}^{+0.08}$ & $ > 116 $ & $ > 110 $ & $ > 129 $ & $1.43_{-0.57}^{+0.74}$ & $1.05_{-0.43}^{+0.52}$ & $1.26_{-0.46}^{+0.57}$ \\
2MASX J19373299-0613046 & 60101003002 & 0761870201 & SW & $2.42_{-0.10}^{+0.16}$ & $2.24_{-0.03}^{+0.04}$ & $2.31_{-0.03}^{+0.04}$ & $ > 129 $ & $180_{-79}^{+552}$ & $ > 115 $ & $1.83_{-0.61}^{+0.74}$ & $1.16_{-0.39}^{+0.57}$ & $1.31_{-0.40}^{+0.56}$ \\
 & 60702018004 & 0891010101 & SW & $2.45_{-0.10}^{+0.11}$ & $2.33_{-0.07}^{+0.07}$ & $2.44_{-0.07}^{+0.07}$ & $ > 220 $ & $ > 202 $ & $ > 264 $ & $1.52_{-0.55}^{+0.85}$ & $1.32_{-0.49}^{+0.80}$ & $1.56_{-0.52}^{+0.89}$ \\
 & 60702018004 & 0891010201 & SW & $2.30_{-0.07}^{+0.09}$ & $2.28_{-0.06}^{+0.07}$ & $2.40_{-0.07}^{+0.07}$ & $ > 125 $ & $ > 120 $ & $ > 140 $ & $0.78_{-0.34}^{+0.43}$ & $1.15_{-0.37}^{+0.42}$ & $1.37_{-0.43}^{+0.45}$ \\
NGC 6814 & 60701012002 & 0885090101 & LW & $1.83_{-0.07}^{+0.07}$ & $1.74_{-0.02}^{+0.02}$ & $1.80_{-0.02}^{+0.04}$ & $ > 484 $ & $ > 274 $ & $ > 353 $ & $0.46_{-0.20}^{+0.24}$ & $0.28_{-0.15}^{+0.17}$ & $0.46_{-0.16}^{+0.19}$ \\
SWIFT J212745.6+565636 & 60001110002 & 0693781701 & SW & $1.87_{-0.05}^{+0.08}$ & $1.82_{-0.04}^{+0.06}$ & $1.93_{-0.06}^{+0.06}$ & $66_{-12}^{+20}$ & $64_{-13}^{+20}$ & $70_{-15}^{+24}$ & $1.39_{-0.35}^{+0.46}$ & $1.18_{-0.31}^{+0.36}$ & $1.50_{-0.36}^{+0.43}$ \\
 & 60001110003 & 0693781701 & SW & $1.94_{-0.12}^{+0.12}$ & $1.89_{-0.07}^{+0.07}$ & $2.00_{-0.07}^{+0.07}$ & $92_{-28}^{+64}$ & $87_{-23}^{+45}$ & $98_{-28}^{+59}$ & $1.43_{-0.45}^{+0.56}$ & $1.36_{-0.37}^{+0.44}$ & $1.65_{-0.41}^{+0.49}$ \\
 & 60001110005 & 0693781801 & SW & $2.00_{-0.07}^{+0.07}$ & $1.88_{-0.04}^{+0.04}$ & $1.99_{-0.04}^{+0.04}$ & $87_{-17}^{+27}$ & $73_{-11}^{+15}$ & $82_{-13}^{+19}$ & $1.73_{-0.32}^{+0.36}$ & $1.47_{-0.24}^{+0.27}$ & $1.80_{-0.28}^{+0.31}$ \\
 & 60001110007 & 0693781901 & SW & $1.96_{-0.08}^{+0.09}$ & $1.90_{-0.05}^{+0.05}$ & $2.01_{-0.05}^{+0.05}$ & $59_{-10}^{+16}$ & $54_{-8}^{+11}$ & $59_{-9}^{+14}$ & $1.71_{-0.38}^{+0.44}$ & $1.57_{-0.34}^{+0.38}$ & $1.92_{-0.36}^{+0.42}$ \\
2MASX J21344509-2725557 & 60363005002 & 0802200201 & LW & $1.87_{-0.18}^{+0.24}$ & $1.77_{-0.08}^{+0.19}$ & $1.84_{-0.08}^{+0.08}$ & $ > 46 $ & $ > 40 $ & $ > 45 $ & $0.75_{-0.74}^{+1.30}$ & $ < 1.54 $ & $0.80_{-0.71}^{+1.20}$ \\
NGC 7314 & 60201031002 & 0790650101 & SW & $2.01_{-0.09}^{+0.09}$ & $1.93_{-0.06}^{+0.06}$ & $2.04_{-0.06}^{+0.06}$ & $184_{-74}^{+321}$ & $165_{-58}^{+180}$ & $200_{-78}^{+308}$ & $1.00_{-0.30}^{+0.36}$ & $0.90_{-0.26}^{+0.32}$ & $1.09_{-0.28}^{+0.34}$ \\
Mrk 915 & 60002060002 & 0744490401 & LW & $1.75_{-0.11}^{+0.09}$ & $1.66_{-0.09}^{+0.09}$ & $1.77_{-0.09}^{+0.09}$ & $ > 363 $ & $ > 290 $ & $ > 322 $ & $ < 0.61 $ & $ < 0.43 $ & $0.32_{-0.24}^{+0.31}$ \\
 & 60002060004 & 0744490501 & LW & $1.40_{-0.23}^{+0.24}$ & $1.41_{-0.16}^{+0.16}$ & $1.52_{-0.16}^{+0.16}$ & $ > 33 $ & $74_{-36}^{+434}$ & $ > 44 $ & $ < 0.41 $ & $ < 0.42 $ & $ < 0.58 $ \\
 & 60002060006 & 0744490601 & LW & $1.75_{-0.48}^{+0.48}$ & $1.55_{-0.23}^{+0.27}$ & $1.67_{-0.25}^{+0.27}$ & $ > 40 $ & $ > 39 $ & $ > 42 $ & $ < 3.43 $ & $ < 1.77 $ & $ < 2.17 $ \\
MR 2251-178 & 60102025002 & 0763920501 & SW & $1.68_{-0.10}^{+0.12}$ & $1.59_{-0.16}^{+0.14}$ & $1.70_{-0.12}^{+0.12}$ & $ > 74 $ & $ > 66 $ & $ > 75 $ & $ < 0.10 $ & $ < 0.08 $ & $ < 0.10 $ \\
 & 60102025004 & 0763920601 & SW & $1.75_{-0.11}^{+0.11}$ & $1.63_{-0.07}^{+0.07}$ & $1.74_{-0.07}^{+0.07}$ & $117_{-43}^{+146}$ & $100_{-32}^{+66}$ & $111_{-37}^{+106}$ & $ < 0.46 $ & $ < 0.25 $ & $ < 0.39 $ \\
 & 60102025006 & 0763920701 & SW & $1.86_{-0.11}^{+0.11}$ & $1.66_{-0.04}^{+0.08}$ & $1.77_{-0.08}^{+0.08}$ & $ > 95 $ & $111_{-44}^{+199}$ & $123_{-50}^{+262}$ & $0.30_{-0.28}^{+0.34}$ & $ < 0.50 $ & $0.25_{-0.25}^{+0.30}$ \\
 & 60102025008 & 0763920801 & SW & $1.85_{-0.16}^{+0.14}$ & $1.63_{-0.07}^{+0.08}$ & $1.74_{-0.08}^{+0.08}$ & $ > 78 $ & $114_{-47}^{+132}$ & $130_{-56}^{+412}$ & $0.32_{-0.31}^{+0.37}$ & $ < 0.28 $ & $ < 0.42 $ \\
 & 90601637002 & 0872390801 & SW & $1.58_{-0.13}^{+0.14}$ & $1.56_{-0.11}^{+0.11}$ & $1.68_{-0.11}^{+0.11}$ & $67_{-23}^{+66}$ & $74_{-26}^{+78}$ & $79_{-28}^{+90}$ & $ < 0.51 $ & $ < 0.43 $ & $ < 0.58 $ \\
NGC 7469 & 60101001002 & 0760350201 & SW & $1.92_{-0.07}^{+0.12}$ & $1.83_{-0.03}^{+0.03}$ & $1.91_{-0.04}^{+0.07}$ & $ > 125 $ & $ > 109 $ & $ > 125 $ & $0.67_{-0.30}^{+0.39}$ & $0.57_{-0.26}^{+0.30}$ & $0.74_{-0.28}^{+0.34}$ \\
 & 60101001004 & 0760350301 & SW & $1.87_{-0.08}^{+0.09}$ & $1.80_{-0.03}^{+0.07}$ & $1.90_{-0.06}^{+0.07}$ & $ > 239 $ & $ > 171 $ & $ > 200 $ & $0.48_{-0.26}^{+0.36}$ & $0.47_{-0.26}^{+0.31}$ & $0.67_{-0.28}^{+0.36}$ \\
 & 60101001006 & 0760350401 & SW & $1.89_{-0.06}^{+0.07}$ & $1.77_{-0.04}^{+0.04}$ & $1.84_{-0.04}^{+0.04}$ & $ > 274 $ & $ > 170 $ & $ > 210 $ & $0.72_{-0.30}^{+0.42}$ & $0.45_{-0.15}^{+0.34}$ & $0.62_{-0.16}^{+0.37}$ \\
 & 60101001010 & 0760350601 & SW & $1.79_{-0.07}^{+0.07}$ & $1.74_{-0.05}^{+0.05}$ & $1.81_{-0.05}^{+0.05}$ & $ > 168 $ & $ > 153 $ & $ > 179 $ & $0.32_{-0.26}^{+0.40}$ & $ < 0.51 $ & $0.33_{-0.22}^{+0.35}$ \\
 & 60101001012 & 0760350701 & SW & $1.89_{-0.08}^{+0.09}$ & $1.78_{-0.05}^{+0.05}$ & $1.85_{-0.05}^{+0.05}$ & $ > 463 $ & $ > 306 $ & $ > 386 $ & $0.96_{-0.40}^{+0.55}$ & $0.64_{-0.27}^{+0.36}$ & $0.88_{-0.32}^{+0.41}$ \\
 & 60101001014 & 0760350801 & SW & $1.86_{-0.06}^{+0.06}$ & $1.74_{-0.04}^{+0.04}$ & $1.82_{-0.04}^{+0.04}$ & $ > 230 $ & $ > 189 $ & $ > 212 $ & $0.43_{-0.23}^{+0.35}$ & $0.23_{-0.18}^{+0.27}$ & $0.37_{-0.19}^{+0.30}$ \\
Mrk 926 & 60201029002 & 0790640101 & SW & $1.78_{-0.06}^{+0.06}$ & $1.70_{-0.05}^{+0.05}$ & $1.77_{-0.05}^{+0.05}$ & $ > 184 $ & $ > 159 $ & $ > 189 $ & $ < 0.36 $ & $ < 0.18 $ & $ < 0.29 $ \\
LCRS B232242.2-384320 & 80502607002 & 0840800501 & SW & $1.75_{-0.09}^{+0.10}$ & $1.71_{-0.07}^{+0.07}$ & $1.78_{-0.07}^{+0.07}$ & $ > 155 $ & $ > 155 $ & $ > 169 $ & $ < 0.71 $ & $ < 0.49 $ & $0.28_{-0.26}^{+0.42}$ \\
RX J1231.6+7044 & 60701055002 & 0891804001 & SW & $1.79_{-0.11}^{+0.14}$ & $1.69_{-0.08}^{+0.09}$ & $1.75_{-0.08}^{+0.11}$ & $ > 122 $ & $ > 91 $ & $ > 133 $ & $ < 0.29 $ & $ < 0.15 $ & $ < 0.19 $ \\
NGC 4579 & 60201051002 & 0790840201 & FF & $1.83_{-0.16}^{+0.18}$ & $1.92_{-0.14}^{+0.17}$ & $1.99_{-0.13}^{+0.16}$ & $ > 60 $ & $ > 56 $ & $ > 58 $ & $ < 0.99 $ & $ < 1.31 $ & $ < 1.70 $ \\
MCG +00-58-028 & 60001147002 & 0743010501 & FF & $2.00_{-0.20}^{+0.27}$ & $1.66_{-0.10}^{+0.10}$ & $1.74_{-0.10}^{+0.10}$ & $ > 97 $ & $ > 29 $ & $ > 33 $ & $1.15_{-0.99}^{+2.27}$ & $ < 1.23 $ & $ < 1.46 $ \\
\enddata 
\tablecomments{
See \S \ref{sec:fitting} for the definition of these parameters. }
\end{deluxetable*}

\startlongtable
 \begin{deluxetable*}{cccccccc}
\tabletypesize{\scriptsize}
\renewcommand\tabcolsep{2.2pt}
\tablecaption{Fitting results using \Nu data alone (without requring simultaneity with joint XMM exposures) \label{tab:sole}}
\tablehead{
\colhead{Source} & \colhead{Nu ID} & \colhead{XMM ID} & \colhead{Nu exp}  & \colhead{Overlap Nu exp} & \colhead{$\Gamma$} &  \colhead{$E_{\rm cut}$} & \colhead{$R$}\\
\colhead{} & \colhead{} & \colhead{} & \colhead{ks} & \colhead{ks} & \colhead{} & \colhead{keV} &  \colhead{}   
}  
\startdata
Fairall 9 & 60001130002 & 0741330101 & 49.2 & 3.0 & $1.87_{-0.04}^{+0.04}$ & $ > 156 $ & $ < 1.05 $ \\
Mrk 359 & 60402021004 & 0830550901 & 49.7 & 23.6 & $1.82_{-0.09}^{+0.10}$ & $ > 109 $ & $0.85_{-0.58}^{+0.97}$ \\
 & 60402021006 & 0830551001 & 51.0 & 26.9 & $1.81_{-0.07}^{+0.14}$ & $ > 205 $ & $0.53_{-0.43}^{+0.72}$ \\
 & 60402021008 & 0830551101 & 48.0 & 29.1 & $1.92_{-0.20}^{+0.19}$ & $ > 85 $ & $ < 1.97 $ \\
NGC 931 & 60101002002 & 0760530201 & 63.0 & 44.6 & $1.88_{-0.07}^{+0.07}$ & $ > 253 $ & $0.73_{-0.23}^{+0.27}$ \\
 & 60101002004 & 0760530301 & 64.2 & 38.7 & $1.88_{-0.06}^{+0.05}$ & $ > 343 $ & $0.80_{-0.25}^{+0.33}$ \\
Mrk 1044 & 60401005002 & 0824080301 & 267.1 & 65.6 & $2.49_{-0.05}^{+0.05}$ & $ > 301 $ & $1.75_{-0.50}^{+0.61}$ \\
3C 109 & 60301011002 & 0795600101 & 67.7 & 28.1 & $1.73_{-0.17}^{+0.20}$ & $ > 83 $ & $ < 0.60 $ \\
NGC 1566 & 80301601002 & 0800840201 & 56.8 & 46.4 & $1.83_{-0.05}^{+0.05}$ & $ > 453 $ & $0.74_{-0.15}^{+0.16}$ \\
 & 80401601002 & 0820530401 & 75.4 & 52.7 & $1.78_{-0.07}^{+0.07}$ & $ > 352 $ & $0.67_{-0.25}^{+0.29}$ \\
 & 80502606002 & 0840800401 & 57.3 & 44.7 & $1.75_{-0.14}^{+0.09}$ & $ > 190 $ & $0.42_{-0.28}^{+0.36}$ \\
1H 0419-577 & 60402006002 & 0820360101 & 64.2 & 31.1 & $1.67_{-0.13}^{+0.13}$ & $76_{-23}^{+55}$ & $ < 0.55 $ \\
 & 60402006004 & 0820360201 & 48.3 & 20.7 & $1.59_{-0.06}^{+0.11}$ & $76_{-21}^{+34}$ & $ < 0.52 $ \\
Ark 120 & 60001044002 & 0693781501 & 55.3 & 42.5 & $1.82_{-0.04}^{+0.07}$ & $ > 557 $ & $0.49_{-0.17}^{+0.27}$ \\
 & 60001044004 & 0721600401 & 65.5 & 52.0 & $1.97_{-0.03}^{+0.03}$ & $ > 588 $ & $0.69_{-0.18}^{+0.18}$ \\
ESO 362-18 & 60201046002 & 0790810101 & 101.9 & 58.2 & $1.55_{-0.08}^{+0.09}$ & $130_{-40}^{+97}$ & $0.56_{-0.31}^{+0.39}$ \\
2MASX J05210136-2521450 & 60201022002 & 0790580101 & 155.1 & 37.0 & $2.13_{-0.09}^{+0.10}$ & $ > 218 $ & $ < 0.87 $ \\
Mrk 79 & 60601010004 & 0870880101 & 38.5 & 13.5 & $1.80_{-0.05}^{+0.05}$ & $201_{-86}^{+523}$ & $0.62_{-0.42}^{+0.61}$ \\
Mrk 110 & 60502022002 & 0852590101 & 86.8 & 10.8 & $1.80_{-0.06}^{+0.06}$ & $188_{-69}^{+246}$ & $ < 0.39 $ \\
 & 60502022004 & 0852590201 & 88.7 & 24.5 & $1.78_{-0.06}^{+0.06}$ & $ > 166 $ & $0.38_{-0.30}^{+0.37}$ \\
NGC 2992 & 90501623002 & 0840920301 & 57.5 & 53.1 & $1.67_{-0.04}^{+0.04}$ & $351_{-129}^{+476}$ & $ < 0.16 $ \\
MCG -05-23-016 & 60701014002 & 0890670101 & 83.7 & 37.7 & $1.85_{-0.03}^{+0.03}$ & $118_{-15}^{+20}$ & $0.77_{-0.15}^{+0.17}$ \\
NGC 3227 & 60202002002 & 0782520201 & 49.8 & 25.7 & $1.73_{-0.07}^{+0.07}$ & $165_{-46}^{+97}$ & $1.20_{-0.33}^{+0.40}$ \\
 & 60202002004 & 0782520301 & 42.5 & 23.4 & $1.61_{-0.04}^{+0.09}$ & $113_{-29}^{+58}$ & $0.88_{-0.33}^{+0.41}$ \\
 & 60202002006 & 0782520401 & 39.7 & 30.3 & $1.83_{-0.08}^{+0.08}$ & $ > 206 $ & $0.94_{-0.27}^{+0.32}$ \\
 & 60202002008 & 0782520501 & 41.8 & 32.7 & $1.89_{-0.03}^{+0.04}$ & $ > 652 $ & $0.89_{-0.23}^{+0.25}$ \\
 & 60202002010 & 0782520601 & 40.9 & 39.1 & $1.80_{-0.05}^{+0.07}$ & $232_{-80}^{+257}$ & $0.94_{-0.23}^{+0.26}$ \\
 & 60202002012 & 0782520701 & 39.3 & 29.5 & $1.92_{-0.07}^{+0.03}$ & $ > 347 $ & $1.06_{-0.28}^{+0.33}$ \\
 & 80502609002 & 0844341301 & 28.8 & 25.7 & $1.67_{-0.10}^{+0.10}$ & $ > 131 $ & $0.33_{-0.21}^{+0.25}$ \\
 & 80502609004 & 0844341401 & 27.7 & 20.4 & $1.50_{-0.19}^{+0.19}$ & $ > 67 $ & $0.68_{-0.51}^{+0.77}$ \\
2MASSi J1031543-141651 & 60701046002 & 0890410101 & 226.9 & 47.9 & $1.80_{-0.03}^{+0.04}$ & $159_{-39}^{+63}$ & $0.70_{-0.33}^{+0.41}$ \\
NGC 3516 & 60160001002 & 0854591101 & 39.9 & 7.8 & $1.96_{-0.07}^{+0.06}$ & $ > 656 $ & $0.77_{-0.50}^{+0.69}$ \\
HE 1136-2304 & 80002031002 & 0741260101 & 23.8 & 23.7 & $1.79_{-0.16}^{+0.14}$ & $ > 92 $ & $ < 0.79 $ \\
 & 80002031003 & 0741260101 & 63.6 & 23.0 & $1.60_{-0.05}^{+0.05}$ & $92_{-29}^{+78}$ & $ < 0.50 $ \\
KUG 1141+371 & 90601618002 & 0871190101 & 38.6 & 9.7 & $1.82_{-0.09}^{+0.14}$ & $ > 73 $ & $0.94_{-0.88}^{+1.50}$ \\
2MASX J11454045-1827149 & 60302002002 & 0795580101 & 21.0 & 16.7 & $1.77_{-0.13}^{+0.13}$ & $85_{-31}^{+105}$ & $0.51_{-0.39}^{+0.51}$ \\
 & 60302002004 & 0795580201 & 20.8 & 12.0 & $1.73_{-0.11}^{+0.14}$ & $ > 72 $ & $ < 0.84 $ \\
 & 60302002006 & 0795580301 & 23.1 & 15.1 & $1.75_{-0.06}^{+0.06}$ & $91_{-26}^{+60}$ & $ < 0.64 $ \\
 & 60302002008 & 0795580401 & 20.7 & 14.3 & $1.77_{-0.10}^{+0.15}$ & $122_{-52}^{+282}$ & $0.43_{-0.39}^{+0.50}$ \\
 & 60302002010 & 0795580501 & 22.4 & 13.8 & $1.79_{-0.11}^{+0.13}$ & $ > 79 $ & $0.50_{-0.49}^{+0.56}$ \\
NGC 4051 & 60401009002 & 0830430201 & 311.1 & 31.5 & $2.04_{-0.03}^{+0.03}$ & $ > 2384 $ & $1.67_{-0.48}^{+0.60}$ \\
NGC 4593 & 60001149002 & 0740920201 & 23.3 & 8.1 & $1.86_{-0.08}^{+0.09}$ & $ > 269 $ & $0.88_{-0.54}^{+0.92}$ \\
 & 60001149004 & 0740920301 & 21.7 & 4.6 & $1.78_{-0.11}^{+0.11}$ & $ > 543 $ & $ < 1.42 $ \\
 & 60001149006 & 0740920401 & 21.3 & 11.2 & $1.72_{-0.07}^{+0.07}$ & $ > 292 $ & $1.18_{-0.56}^{+0.89}$ \\
 & 60001149008 & 0740920501 & 23.1 & 8.0 & $1.82_{-0.05}^{+0.05}$ & $ > 1291 $ & $0.67_{-0.35}^{+0.49}$ \\
 & 60001149010 & 0740920601 & 21.2 & 12.3 & $1.83_{-0.08}^{+0.09}$ & $ > 186 $ & $0.40_{-0.31}^{+0.46}$ \\
MCG -06-30-015 & 60001047003 & 0693781301 & 127.2 & 64.1 & $2.27_{-0.04}^{+0.02}$ & $ > 736 $ & $1.75_{-0.31}^{+0.35}$ \\
 & 60001047005 & 0693781401 & 29.6 & 15.1 & $2.15_{-0.11}^{+0.04}$ & $ > 295 $ & $1.87_{-0.66}^{+0.88}$ \\
IC 4329A & 60702050002 & 0862090101 & 20.6 & 6.7 & $1.77_{-0.06}^{+0.06}$ & $ > 201 $ & $0.58_{-0.27}^{+0.33}$ \\
 & 60702050004 & 0862090301 & 20.1 & 6.2 & $1.70_{-0.07}^{+0.07}$ & $ > 206 $ & $ < 0.54 $ \\
 & 60702050006 & 0862090501 & 19.8 & 11.0 & $1.83_{-0.07}^{+0.02}$ & $ > 404 $ & $0.26_{-0.17}^{+0.20}$ \\
 & 60702050010 & 0862090901 & 18.0 & 1.2 & $1.78_{-0.07}^{+0.07}$ & $ > 171 $ & $ < 2.46 $ \\
NGC 5548 & 60002044003 & 0720110601 & 27.3 & 27.2 & $1.77_{-0.08}^{+0.08}$ & $ > 192 $ & $0.64_{-0.22}^{+0.26}$ \\
 & 60002044005 & 0720111001 & 49.5 & 28.7 & $1.55_{-0.08}^{+0.03}$ & $86_{-17}^{+20}$ & $0.46_{-0.20}^{+0.22}$ \\
 & 60002044008 & 0720111501 & 50.1 & 24.5 & $1.49_{-0.07}^{+0.07}$ & $89_{-17}^{+26}$ & $0.38_{-0.21}^{+0.24}$ \\
ESO 511-G030 & 60502035002 & 0852010101 & 32.1 & 16.5 & $1.70_{-0.26}^{+0.09}$ & $ > 40 $ & $ < 1.51 $ \\
 & 60502035004 & 0852010201 & 34.1 & 18.0 & $1.71_{-0.27}^{+0.20}$ & $ > 48 $ & $ < 0.37 $ \\
 & 60502035006 & 0852010301 & 31.2 & 17.4 & $1.56_{-0.11}^{+0.14}$ & $ > 57 $ & $ < 0.29 $ \\
 & 60502035008 & 0852010401 & 41.8 & 19.8 & $1.64_{-0.10}^{+0.09}$ & $ > 76 $ & $ < 1.23 $ \\
Mrk 1383 & 60501049002 & 0852210101 & 96.0 & 38.0 & $1.88_{-0.07}^{+0.07}$ & $ > 142 $ & $0.55_{-0.42}^{+0.63}$ \\
Mrk 817 & 60702008002 & 0882340601 & 65.0 & 57.8 & $2.16_{-0.08}^{+0.06}$ & $ > 1150 $ & $1.57_{-0.44}^{+0.54}$ \\
 & 60702008004 & 0882340701 & 71.7 & 52.4 & $1.74_{-0.14}^{+0.14}$ & $120_{-47}^{+176}$ & $0.95_{-0.45}^{+0.60}$ \\
 & 60702008006 & 0882340801 & 78.0 & 52.1 & $1.95_{-0.16}^{+0.09}$ & $ > 267 $ & $1.65_{-0.42}^{+0.91}$ \\
Mrk 841 & 60101023002 & 0763790501 & 23.4 & 8.9 & $1.83_{-0.10}^{+0.12}$ & $ > 127 $ & $0.99_{-0.70}^{+1.28}$ \\
 & 60702007002 & 0882130301 & 179.6 & 59.5 & $1.87_{-0.07}^{+0.07}$ & $275_{-116}^{+577}$ & $0.95_{-0.41}^{+0.51}$ \\
 & 80701616002 & 0890640201 & 53.1 & 28.0 & $1.95_{-0.12}^{+0.11}$ & $ > 158 $ & $1.19_{-0.57}^{+0.79}$ \\
3C 382 & 60202015002 & 0790600101 & 23.1 & 9.5 & $1.75_{-0.08}^{+0.07}$ & $ > 133 $ & $ < 0.63 $ \\
 & 60202015004 & 0790600201 & 24.6 & 12.2 & $1.70_{-0.04}^{+0.04}$ & $115_{-31}^{+67}$ & $ < 0.58 $ \\
 & 60202015006 & 0790600301 & 20.8 & 9.0 & $1.74_{-0.04}^{+0.04}$ & $350_{-196}^{+455}$ & $ < 0.27 $ \\
 & 60202015008 & 0790600401 & 21.7 & 10.6 & $1.69_{-0.07}^{+0.10}$ & $168_{-73}^{+489}$ & $ < 0.37 $ \\
 & 60202015010 & 0790600501 & 21.1 & 8.3 & $1.77_{-0.04}^{+0.03}$ & $ > 253 $ & $ < 0.26 $ \\
Fairall 49 & 60301028002 & 0795690101 & 97.8 & 27.2 & $2.42_{-0.07}^{+0.07}$ & $182_{-78}^{+442}$ & $1.43_{-0.57}^{+0.74}$ \\
2MASX J19373299-0613046 & 60101003002 & 0761870201 & 65.5 & 64.0 & $2.41_{-0.10}^{+0.10}$ & $ > 118 $ & $1.83_{-0.61}^{+0.74}$ \\
 & 60702018004 & 0891010201 & 150.0 & 55.1 & $2.38_{-0.06}^{+0.06}$ & $ > 224 $ & $0.78_{-0.34}^{+0.43}$ \\
NGC 6814 & 60701012002 & 0885090101 & 128.2 & 42.0 & $1.83_{-0.04}^{+0.03}$ & $ > 644 $ & $0.46_{-0.20}^{+0.24}$ \\
SWIFT J212745.6+565636 & 60001110002 & 0693781701 & 49.2 & 46.6 & $1.88_{-0.06}^{+0.08}$ & $61_{-10}^{+15}$ & $1.39_{-0.35}^{+0.46}$ \\
 & 60001110003 & 0693781701 & 28.8 & 26.5 & $1.96_{-0.12}^{+0.12}$ & $94_{-28}^{+63}$ & $1.43_{-0.45}^{+0.56}$ \\
 & 60001110005 & 0693781801 & 74.6 & 67.0 & $2.00_{-0.07}^{+0.07}$ & $87_{-16}^{+25}$ & $1.73_{-0.32}^{+0.36}$ \\
 & 60001110007 & 0693781901 & 42.1 & 38.4 & $1.96_{-0.09}^{+0.08}$ & $59_{-9}^{+19}$ & $1.71_{-0.38}^{+0.44}$ \\
2MASX J21344509-2725557 & 60363005002 & 0802200201 & 21.1 & 12.3 & $1.74_{-0.13}^{+0.28}$ & $ > 45 $ & $0.75_{-0.74}^{+1.30}$ \\
NGC 7314 & 60201031002 & 0790650101 & 100.4 & 30.1 & $2.05_{-0.05}^{+0.05}$ & $345_{-138}^{+580}$ & $1.00_{-0.30}^{+0.36}$ \\
Mrk 915 & 60002060002 & 0744490401 & 53.0 & 36.9 & $1.80_{-0.09}^{+0.12}$ & $ > 365 $ & $ < 0.61 $ \\
 & 60002060004 & 0744490501 & 54.2 & 18.1 & $1.54_{-0.13}^{+0.13}$ & $ > 69 $ & $ < 0.62 $ \\
 & 60002060006 & 0744490601 & 50.7 & 7.8 & $1.62_{-0.20}^{+0.19}$ & $ > 65 $ & $ < 3.43 $ \\
MR 2251-178 & 60102025002 & 0763920501 & 23.1 & 5.6 & $1.67_{-0.07}^{+0.06}$ & $137_{-53}^{+148}$ & $ < 0.10 $ \\
 & 60102025004 & 0763920601 & 23.2 & 12.7 & $1.75_{-0.07}^{+0.07}$ & $175_{-66}^{+263}$ & $ < 0.46 $ \\
 & 60102025006 & 0763920701 & 20.6 & 10.6 & $1.82_{-0.08}^{+0.08}$ & $128_{-43}^{+118}$ & $0.30_{-0.28}^{+0.34}$ \\
 & 60102025008 & 0763920801 & 21.7 & 10.2 & $1.79_{-0.10}^{+0.10}$ & $171_{-71}^{+437}$ & $0.32_{-0.31}^{+0.37}$ \\
 & 90601637002 & 0872390801 & 23.6 & 15.9 & $1.63_{-0.10}^{+0.11}$ & $106_{-42}^{+174}$ & $ < 0.51 $ \\
NGC 7469 & 60101001002 & 0760350201 & 21.6 & 21.4 & $1.92_{-0.08}^{+0.13}$ & $ > 119 $ & $0.67_{-0.30}^{+0.39}$ \\
 & 60101001004 & 0760350301 & 20.0 & 19.9 & $1.86_{-0.08}^{+0.09}$ & $ > 222 $ & $0.48_{-0.26}^{+0.36}$ \\
 & 60101001006 & 0760350401 & 22.5 & 16.8 & $1.88_{-0.05}^{+0.06}$ & $ > 312 $ & $0.72_{-0.30}^{+0.42}$ \\
 & 60101001010 & 0760350601 & 20.9 & 10.7 & $1.84_{-0.05}^{+0.05}$ & $ > 304 $ & $0.32_{-0.26}^{+0.40}$ \\
 & 60101001012 & 0760350701 & 21.0 & 10.5 & $1.84_{-0.05}^{+0.05}$ & $ > 768 $ & $0.96_{-0.40}^{+0.55}$ \\
 & 60101001014 & 0760350801 & 23.4 & 15.8 & $1.83_{-0.05}^{+0.05}$ & $ > 201 $ & $0.43_{-0.23}^{+0.35}$ \\
Mrk 926 & 60201029002 & 0790640101 & 106.2 & 6.1 & $1.72_{-0.02}^{+0.02}$ & $296_{-86}^{+196}$ & $ < 0.36 $ \\
LCRS B232242.2-384320 & 80502607002 & 0840800501 & 54.8 & 38.6 & $1.74_{-0.07}^{+0.08}$ & $ > 196 $ & $ < 0.71 $ \\
RX J1231.6+7044 & 60701055002 & 0891804001 & 136.5 & 18.2 & $1.77_{-0.06}^{+0.04}$ & $ > 229 $ & $ < 0.29 $ \\
NGC 4579 & 60201051002 & 0790840201 & 117.8 & 8.1 & $1.81_{-0.05}^{+0.04}$ & $ > 125 $ & $ < 0.09 $ \\
MCG +00-58-028 & 60001147002 & 0743010501 & 26.7 & 8.3 & $1.96_{-0.12}^{+0.15}$ & $ > 232 $ & $1.15_{-0.99}^{+2.27}$ \\
\enddata 
\tablecomments{``Nu exp" refers to the net exposure time of \Nu FPMA using all available GTI, while ``Overlap Nu exp" refers to the remaining net exposure after combining the \Nu GTI with the EPIC-pn GTI of the joint XMM observation.  
The fitting results in this table are derived with all the available \Nu data (``Nu exp"), not requiring simultaneity with corresponding EPIC-pn exposure.}
\end{deluxetable*}

\bibliography{NuXMM}{}

\begin{thebibliography}{}
\expandafter\ifx\csname natexlab\endcsname\relax\def\natexlab#1{#1}\fi
\providecommand{\url}[1]{\href{#1}{#1}}
\providecommand{\dodoi}[1]{doi:~\href{http://doi.org/#1}{\nolinkurl{#1}}}
\providecommand{\doeprint}[1]{\href{http://ascl.net/#1}{\nolinkurl{http://ascl.net/#1}}}
\providecommand{\doarXiv}[1]{\href{https://arxiv.org/abs/#1}{\nolinkurl{https://arxiv.org/abs/#1}}}

\bibitem[{{Akylas} \& {Georgantopoulos}(2021)}]{Akylas_2021}
{Akylas}, A., \& {Georgantopoulos}, I. 2021, \aap, 655, A60,
  \dodoi{10.1051/0004-6361/202141186}

\bibitem[{{Arnaud}(1996)}]{Arnaud_1996}
{Arnaud}, K.~A. 1996, in Astronomical Society of the Pacific Conference Series,
  Vol. 101, Astronomical Data Analysis Software and Systems V, ed. G.~H.
  {Jacoby} \& J.~{Barnes}, 17

\bibitem[{{Balokovi{\'c}} {et~al.}(2020){Balokovi{\'c}}, {Harrison},
  {Madejski}, {Comastri}, {Ricci}, {Annuar}, {Ballantyne}, {Boorman}, {Brandt},
  {Brightman}, {Gandhi}, {Kamraj}, {Koss}, {Marchesi}, {Marinucci}, {Masini},
  {Matt}, {Stern}, \& {Urry}}]{Balokovi_2020}
{Balokovi{\'c}}, M., {Harrison}, F.~A., {Madejski}, G., {et~al.} 2020, \apj,
  905, 41, \dodoi{10.3847/1538-4357/abc342}

\bibitem[{{Brenneman} {et~al.}(2014){Brenneman}, {Madejski}, {Fuerst}, {Matt},
  {Elvis}, {Harrison}, {Ballantyne}, {Boggs}, {Christensen}, {Craig}, {Fabian},
  {Grefenstette}, {Hailey}, {Madsen}, {Marinucci}, {Rivers}, {Stern}, {Walton},
  \& {Zhang}}]{Brenneman_2014}
{Brenneman}, L.~W., {Madejski}, G., {Fuerst}, F., {et~al.} 2014, \apj, 788, 61,
  \dodoi{10.1088/0004-637X/788/1/61}

\bibitem[{{Cappi} {et~al.}(2016){Cappi}, {De Marco}, {Ponti}, {Ursini},
  {Petrucci}, {Bianchi}, {Kaastra}, {Kriss}, {Mehdipour}, {Whewell}, {Arav},
  {Behar}, {Boissay}, {Branduardi-Raymont}, {Costantini}, {Ebrero}, {Di Gesu},
  {Harrison}, {Kaspi}, {Matt}, {Paltani}, {Peterson}, {Steenbrugge}, \&
  {Walton}}]{gamma_xmm_1}
{Cappi}, M., {De Marco}, B., {Ponti}, G., {et~al.} 2016, \aap, 592, A27,
  \dodoi{10.1051/0004-6361/201628464}

\bibitem[{{Carter} \& {Read}(2007)}]{Carter_2007}
{Carter}, J.~A., \& {Read}, A.~M. 2007, \aap, 464, 1155,
  \dodoi{10.1051/0004-6361:20065882}

\bibitem[{{Chiang} {et~al.}(2000){Chiang}, {Reynolds}, {Blaes}, {Nowak},
  {Murray}, {Madejski}, {Marshall}, \& {Magdziarz}}]{Chiang_2000}
{Chiang}, J., {Reynolds}, C.~S., {Blaes}, O.~M., {et~al.} 2000, \apj, 528, 292,
  \dodoi{10.1086/308178}

\bibitem[{{Dauser} {et~al.}(2010){Dauser}, {Wilms}, {Reynolds}, \&
  {Brenneman}}]{Dauser_2010}
{Dauser}, T., {Wilms}, J., {Reynolds}, C.~S., \& {Brenneman}, L.~W. 2010,
  \mnras, 409, 1534, \dodoi{10.1111/j.1365-2966.2010.17393.x}

\bibitem[{{De Luca} \& {Molendi}(2004)}]{DeLuca_2004}
{De Luca}, A., \& {Molendi}, S. 2004, \aap, 419, 837,
  \dodoi{10.1051/0004-6361:20034421}

\bibitem[{{Fabian} {et~al.}(2015){Fabian}, {Lohfink}, {Kara}, {Parker},
  {Vasudevan}, \& {Reynolds}}]{Fabian_2015}
{Fabian}, A.~C., {Lohfink}, A., {Kara}, E., {et~al.} 2015, \mnras, 451, 4375,
  \dodoi{10.1093/mnras/stv1218}

\bibitem[{{Feigelson} \& {Nelson}(1985)}]{Feigelson_1985}
{Feigelson}, E.~D., \& {Nelson}, P.~I. 1985, \apj, 293, 192,
  \dodoi{10.1086/163225}

\bibitem[{{Garc{\'\i}a} {et~al.}(2014){Garc{\'\i}a}, {Dauser}, {Lohfink},
  {Kallman}, {Steiner}, {McClintock}, {Brenneman}, {Wilms}, {Eikmann},
  {Reynolds}, \& {Tombesi}}]{Garcia_2014}
{Garc{\'\i}a}, J., {Dauser}, T., {Lohfink}, A., {et~al.} 2014, \apj, 782, 76,
  \dodoi{10.1088/0004-637X/782/2/76}

\bibitem[{{Haardt} \& {Maraschi}(1991)}]{Haardt_1991}
{Haardt}, F., \& {Maraschi}, L. 1991, \apjl, 380, L51, \dodoi{10.1086/186171}

\bibitem[{{Haardt} \& {Maraschi}(1993)}]{Haardt_1993}
---. 1993, \apj, 413, 507, \dodoi{10.1086/173020}

\bibitem[{Harrison {et~al.}(2013)Harrison, Craig, Christensen, Hailey, Zhang,
  Boggs, Stern, Cook, Forster, Giommi, Grefenstette, Kim, Kitaguchi, Koglin,
  Madsen, Mao, Miyasaka, Mori, Perri, Pivovaroff, Puccetti, Rana, Westergaard,
  Willis, Zoglauer, An, Bachetti, Barri{\`{e}}re, Bellm, Bhalerao, Brejnholt,
  Fuerst, Liebe, Markwardt, Nynka, Vogel, Walton, Wik, Alexander, Cominsky,
  Hornschemeier, Hornstrup, Kaspi, Madejski, Matt, Molendi, Smith, Tomsick,
  Ajello, Ballantyne, Balokovi{\'{c}}, Barret, Bauer, Blandford, Brandt,
  Brenneman, Chiang, Chakrabarty, Chenevez, Comastri, Dufour, Elvis, Fabian,
  Farrah, Fryer, Gotthelf, Grindlay, Helfand, Krivonos, Meier, Miller,
  Natalucci, Ogle, Ofek, Ptak, Reynolds, Rigby, Tagliaferri, Thorsett,
  Treister, \& Urry}]{Harrison_2013}
Harrison, F.~A., Craig, W.~W., Christensen, F.~E., {et~al.} 2013, The
  Astrophysical Journal, 770, 103, \dodoi{10.1088/0004-637x/770/2/103}

\bibitem[{{Hinkle} \& {Mushotzky}(2021)}]{Hinkle_2021}
{Hinkle}, J.~T., \& {Mushotzky}, R. 2021, \mnras, 506, 4960,
  \dodoi{10.1093/mnras/stab1976}

\bibitem[{{Jansen} {et~al.}(2001){Jansen}, {Lumb}, {Altieri}, {Clavel}, {Ehle},
  {Erd}, {Gabriel}, {Guainazzi}, {Gondoin}, {Much}, {Munoz}, {Santos},
  {Schartel}, {Texier}, \& {Vacanti}}]{Jansen_2001_XMM}
{Jansen}, F., {Lumb}, D., {Altieri}, B., {et~al.} 2001, \aap, 365, L1,
  \dodoi{10.1051/0004-6361:20000036}

\bibitem[{{Jiang} {et~al.}(2022){Jiang}, {Abdikamalov}, {Bambi}, \&
  {Reynolds}}]{Jiang_2022}
{Jiang}, J., {Abdikamalov}, A.~B., {Bambi}, C., \& {Reynolds}, C.~S. 2022,
  \mnras, 514, 3246, \dodoi{10.1093/mnras/stac1369}

\bibitem[{Kamraj {et~al.}(2018)Kamraj, Harrison, Balokovi{\'{c}}, Lohfink, \&
  Brightman}]{Kamraj_2018}
Kamraj, N., Harrison, F.~A., Balokovi{\'{c}}, M., Lohfink, A., \& Brightman, M.
  2018, The Astrophysical Journal, 866, 124, \dodoi{10.3847/1538-4357/aadd0d}

\bibitem[{{Kamraj} {et~al.}(2022){Kamraj}, {Brightman}, {Harrison}, {Stern},
  {Garc{\'\i}a}, {Balokovi{\'c}}, {Ricci}, {Koss}, {Mej{\'\i}a-Restrepo}, {Oh},
  {Powell}, \& {Urry}}]{Kamraj_2022}
{Kamraj}, N., {Brightman}, M., {Harrison}, F.~A., {et~al.} 2022, \apj, 927, 42,
  \dodoi{10.3847/1538-4357/ac45f6}

\bibitem[{{Kang} {et~al.}(2020){Kang}, {Wang}, \& {Kang}}]{Kang_2020}
{Kang}, J., {Wang}, J., \& {Kang}, W. 2020, \apj, 901, 111,
  \dodoi{10.3847/1538-4357/abadf5}

\bibitem[{{Kang} \& {Wang}(2022)}]{Kang_2022}
{Kang}, J.-L., \& {Wang}, J.-X. 2022, \apj, 929, 141,
  \dodoi{10.3847/1538-4357/ac5d49}

\bibitem[{{Kang} \& {Wang}(2023)}]{Kang_2023}
---. 2023, \mnras, 519, 3635, \dodoi{10.1093/mnras/stac3598}

\bibitem[{{Kang} {et~al.}(2021){Kang}, {Wang}, \& {Kang}}]{Kang_2021}
{Kang}, J.-L., {Wang}, J.-X., \& {Kang}, W.-Y. 2021, \mnras, 502, 80,
  \dodoi{10.1093/mnras/stab039}

\bibitem[{{Kara} {et~al.}(2015){Kara}, {Zoghbi}, {Marinucci}, {Walton},
  {Fabian}, {Risaliti}, {Boggs}, {Christensen}, {Fuerst}, {Hailey}, {Harrison},
  {Matt}, {Parker}, {Reynolds}, {Stern}, \& {Zhang}}]{Kara_2015}
{Kara}, E., {Zoghbi}, A., {Marinucci}, A., {et~al.} 2015, \mnras, 446, 737,
  \dodoi{10.1093/mnras/stu2136}

\bibitem[{{Kuntz} \& {Snowden}(2008)}]{Kuntz_2008}
{Kuntz}, K.~D., \& {Snowden}, S.~L. 2008, \aap, 478, 575,
  \dodoi{10.1051/0004-6361:20077912}

\bibitem[{{Lanzuisi} {et~al.}(2016){Lanzuisi}, {Perna}, {Comastri}, {Cappi},
  {Dadina}, {Marinucci}, {Masini}, {Matt}, {Vagnetti}, {Vignali}, {Ballantyne},
  {Bauer}, {Boggs}, {Brandt}, {Brusa}, {Christensen}, {Craig}, {Fabian},
  {Farrah}, {Hailey}, {Harrison}, {Luo}, {Piconcelli}, {Puccetti}, {Ricci},
  {Saez}, {Stern}, {Walton}, \& {Zhang}}]{Lanzuisi_2016}
{Lanzuisi}, G., {Perna}, M., {Comastri}, A., {et~al.} 2016, \aap, 590, A77,
  \dodoi{10.1051/0004-6361/201628325}

\bibitem[{{Lanzuisi} {et~al.}(2019){Lanzuisi}, {Gilli}, {Cappi}, {Dadina},
  {Bianchi}, {Brusa}, {Chartas}, {Civano}, {Comastri}, {Marinucci}, {Middei},
  {Piconcelli}, {Vignali}, {Brandt}, {Tombesi}, \& {Gaspari}}]{Lanzuisi_2019}
{Lanzuisi}, G., {Gilli}, R., {Cappi}, M., {et~al.} 2019, \apjl, 875, L20,
  \dodoi{10.3847/2041-8213/ab15dc}

\bibitem[{{Madsen} {et~al.}(2015){Madsen}, {Harrison}, {Markwardt}, {An},
  {Grefenstette}, {Bachetti}, {Miyasaka}, {Kitaguchi}, {Bhalerao}, {Boggs},
  {Christensen}, {Craig}, {Forster}, {Fuerst}, {Hailey}, {Perri}, {Puccetti},
  {Rana}, {Stern}, {Walton}, {J{\o}rgen Westergaard}, \& {Zhang}}]{Madsen_2015}
{Madsen}, K.~K., {Harrison}, F.~A., {Markwardt}, C.~B., {et~al.} 2015, \apjs,
  220, 8, \dodoi{10.1088/0067-0049/220/1/8}

\bibitem[{{Magdziarz} \& {Zdziarski}(1995)}]{Magdziarz_1995}
{Magdziarz}, P., \& {Zdziarski}, A.~A. 1995, \mnras, 273, 837,
  \dodoi{10.1093/mnras/273.3.837}

\bibitem[{{Mantovani} {et~al.}(2016){Mantovani}, {Nandra}, \&
  {Ponti}}]{Mantovani_2016}
{Mantovani}, G., {Nandra}, K., \& {Ponti}, G. 2016, \mnras, 458, 4198,
  \dodoi{10.1093/mnras/stw596}

\bibitem[{{Marchesi} {et~al.}(2022){Marchesi}, {Zhao}, {Torres-Alb{\`a}},
  {Ajello}, {Gaspari}, {Pizzetti}, {Buchner}, {Bertola}, {Comastri}, {Feltre},
  {Gilli}, {Lanzuisi}, {Matzeu}, {Pozzi}, {Salvestrini}, {Sengupta}, {Silver},
  {Tombesi}, {Traina}, {Vignali}, \& {Zappacosta}}]{Marchesi_2022}
{Marchesi}, S., {Zhao}, X., {Torres-Alb{\`a}}, N., {et~al.} 2022, \apj, 935,
  114, \dodoi{10.3847/1538-4357/ac80be}

\bibitem[{{Marinucci} {et~al.}(2014){Marinucci}, {Matt}, {Kara}, {Miniutti},
  {Elvis}, {Arevalo}, {Ballantyne}, {Balokovi{\'c}}, {Bauer}, {Brenneman},
  {Boggs}, {Cappi}, {Christensen}, {Craig}, {Fabian}, {Fuerst}, {Hailey},
  {Harrison}, {Risaliti}, {Reynolds}, {Stern}, {Walton}, \&
  {Zhang}}]{Marinucci_2014}
{Marinucci}, A., {Matt}, G., {Kara}, E., {et~al.} 2014, \mnras, 440, 2347,
  \dodoi{10.1093/mnras/stu404}

\bibitem[{{Matt} {et~al.}(2015){Matt}, {Balokovi{\'c}}, {Marinucci},
  {Ballantyne}, {Boggs}, {Christensen}, {Comastri}, {Craig}, {Gandhi},
  {Hailey}, {Harrison}, {Madejski}, {Madsen}, {Stern}, \& {Zhang}}]{Matt_2015}
{Matt}, G., {Balokovi{\'c}}, M., {Marinucci}, A., {et~al.} 2015, \mnras, 447,
  3029, \dodoi{10.1093/mnras/stu2653}

\bibitem[{{Middei} {et~al.}(2019){Middei}, {Bianchi}, {Petrucci}, {Ursini},
  {Cappi}, {De Marco}, {De Rosa}, {Malzac}, {Marinucci}, {Matt}, {Ponti}, \&
  {Tortosa}}]{gamma_xmm_2}
{Middei}, R., {Bianchi}, S., {Petrucci}, P.~O., {et~al.} 2019, \mnras, 483,
  4695, \dodoi{10.1093/mnras/sty3379}

\bibitem[{Molina {et~al.}(2019)Molina, Malizia, Bassani, Ursini, Bazzano, \&
  Ubertini}]{Molina_2019}
Molina, M., Malizia, A., Bassani, L., {et~al.} 2019, Monthly Notices of the
  Royal Astronomical Society, 484, 2735, \dodoi{10.1093/mnras/stz156}

\bibitem[{{Nandra} {et~al.}(2007){Nandra}, {O'Neill}, {George}, \&
  {Reeves}}]{Nandra_2007}
{Nandra}, K., {O'Neill}, P.~M., {George}, I.~M., \& {Reeves}, J.~N. 2007,
  \mnras, 382, 194, \dodoi{10.1111/j.1365-2966.2007.12331.x}

\bibitem[{{Oh} {et~al.}(2018){Oh}, {Koss}, {Markwardt}, {Schawinski},
  {Baumgartner}, {Barthelmy}, {Cenko}, {Gehrels}, {Mushotzky}, {Petulante},
  {Ricci}, {Lien}, \& {Trakhtenbrot}}]{Oh_2018}
{Oh}, K., {Koss}, M., {Markwardt}, C.~B., {et~al.} 2018, \apjs, 235, 4,
  \dodoi{10.3847/1538-4365/aaa7fd}

\bibitem[{{Pal} \& {Stalin}(2023)}]{Pal_2023}
{Pal}, I., \& {Stalin}, C.~S. 2023, \mnras, 518, 2529,
  \dodoi{10.1093/mnras/stac3254}

\bibitem[{{Pal} {et~al.}(2022){Pal}, {Stalin}, {Parker}, {Agrawal}, \&
  {Marchesi}}]{Pal_2022}
{Pal}, I., {Stalin}, C.~S., {Parker}, M.~L., {Agrawal}, V.~K., \& {Marchesi},
  S. 2022, \mnras, 517, 3341, \dodoi{10.1093/mnras/stac2736}

\bibitem[{{Panagiotou} \& {Walter}(2019)}]{Panagiotou_2019}
{Panagiotou}, C., \& {Walter}, R. 2019, \aap, 626, A40,
  \dodoi{10.1051/0004-6361/201935052}

\bibitem[{{Panagiotou} \& {Walter}(2020)}]{Panagiotou_2020}
---. 2020, \aap, 640, A31, \dodoi{10.1051/0004-6361/201937390}

\bibitem[{{Parker} {et~al.}(2014){Parker}, {Wilkins}, {Fabian}, {Grupe},
  {Dauser}, {Matt}, {Harrison}, {Brenneman}, {Boggs}, {Christensen}, {Craig},
  {Gallo}, {Hailey}, {Kara}, {Komossa}, {Marinucci}, {Miller}, {Risaliti},
  {Stern}, {Walton}, \& {Zhang}}]{Parker_2014}
{Parker}, M.~L., {Wilkins}, D.~R., {Fabian}, A.~C., {et~al.} 2014, \mnras, 443,
  1723, \dodoi{10.1093/mnras/stu1246}

\bibitem[{{Ponti} {et~al.}(2018){Ponti}, {Bianchi}, {Mu{\~n}oz-Darias}, {Mori},
  {De}, {Rau}, {De Marco}, {Hailey}, {Tomsick}, {Madsen}, {Clavel}, {Rahoui},
  {Lal}, {Roy}, \& {Stern}}]{gamma_xmm_3}
{Ponti}, G., {Bianchi}, S., {Mu{\~n}oz-Darias}, T., {et~al.} 2018, \mnras, 473,
  2304, \dodoi{10.1093/mnras/stx2425}

\bibitem[{{Porquet} {et~al.}(2019){Porquet}, {Done}, {Reeves}, {Grosso},
  {Marinucci}, {Matt}, {Lobban}, {Nardini}, {Braito}, {Marin}, {Kubota},
  {Ricci}, {Koss}, {Stern}, {Ballantyne}, \& {Farrah}}]{Porquet_2019}
{Porquet}, D., {Done}, C., {Reeves}, J.~N., {et~al.} 2019, \aap, 623, A11,
  \dodoi{10.1051/0004-6361/201834448}

\bibitem[{Rani {et~al.}(2019)Rani, Stalin, \& Goswami}]{Rani_2019}
Rani, P., Stalin, C.~S., \& Goswami, K.~D. 2019, Monthly Notices of the Royal
  Astronomical Society, 484, 5113, \dodoi{10.1093/mnras/stz275}

\bibitem[{{Read} \& {Ponman}(2003)}]{Read_2003}
{Read}, A.~M., \& {Ponman}, T.~J. 2003, \aap, 409, 395,
  \dodoi{10.1051/0004-6361:20031099}

\bibitem[{{Risaliti} {et~al.}(2013){Risaliti}, {Harrison}, {Madsen}, {Walton},
  {Boggs}, {Christensen}, {Craig}, {Grefenstette}, {Hailey}, {Nardini},
  {Stern}, \& {Zhang}}]{Risaliti_2013}
{Risaliti}, G., {Harrison}, F.~A., {Madsen}, K.~K., {et~al.} 2013, \nat, 494,
  449, \dodoi{10.1038/nature11938}

\bibitem[{{Sengupta} {et~al.}(2023){Sengupta}, {Marchesi}, {Vignali},
  {Torres-Alb{\`a}}, {Bertola}, {Pizzetti}, {Lanzuisi}, {Salvestrini}, {Zhao},
  {Gaspari}, {Gilli}, {Comastri}, {Traina}, {Tombesi}, {Silver}, {Pozzi}, \&
  {Ajello}}]{Sengupta_2023}
{Sengupta}, D., {Marchesi}, S., {Vignali}, C., {et~al.} 2023, \aap, 676, A103,
  \dodoi{10.1051/0004-6361/202245646}

\bibitem[{Shu {et~al.}(2010)Shu, Yaqoob, \& Wang}]{Shu_2010}
Shu, X.~W., Yaqoob, T., \& Wang, J.~X. 2010, The Astrophysical Journal
  Supplement Series, 187, 581, \dodoi{10.1088/0067-0049/187/2/581}

\bibitem[{{Silver} {et~al.}(2022){Silver}, {Torres-Alb{\`a}}, {Zhao},
  {Marchesi}, {Pizzetti}, {Cox}, {Ajello}, {Cusumano}, {La Parola}, \&
  {Segreto}}]{Silver_2022}
{Silver}, R., {Torres-Alb{\`a}}, N., {Zhao}, X., {et~al.} 2022, \apj, 940, 148,
  \dodoi{10.3847/1538-4357/ac9bf8}

\bibitem[{{Str{\"u}der} {et~al.}(2001){Str{\"u}der}, {Briel}, {Dennerl},
  {Hartmann}, {Kendziorra}, {Meidinger}, {Pfeffermann}, {Reppin}, {Aschenbach},
  {Bornemann}, {Br{\"a}uninger}, {Burkert}, {Elender}, {Freyberg}, {Haberl},
  {Hartner}, {Heuschmann}, {Hippmann}, {Kastelic}, {Kemmer}, {Kettenring},
  {Kink}, {Krause}, {M{\"u}ller}, {Oppitz}, {Pietsch}, {Popp}, {Predehl},
  {Read}, {Stephan}, {St{\"o}tter}, {Tr{\"u}mper}, {Holl}, {Kemmer}, {Soltau},
  {St{\"o}tter}, {Weber}, {Weichert}, {von Zanthier}, {Carathanassis}, {Lutz},
  {Richter}, {Solc}, {B{\"o}ttcher}, {Kuster}, {Staubert}, {Abbey}, {Holland},
  {Turner}, {Balasini}, {Bignami}, {La Palombara}, {Villa}, {Buttler},
  {Gianini}, {Lain{\'e}}, {Lumb}, \& {Dhez}}]{Struder_2001_PN}
{Str{\"u}der}, L., {Briel}, U., {Dennerl}, K., {et~al.} 2001, \aap, 365, L18,
  \dodoi{10.1051/0004-6361:20000066}

\bibitem[{{Tortosa} {et~al.}(2018){Tortosa}, {Bianchi}, {Marinucci}, {Matt}, \&
  {Petrucci}}]{Tortosa_2018}
{Tortosa}, A., {Bianchi}, S., {Marinucci}, A., {Matt}, G., \& {Petrucci}, P.~O.
  2018, \aap, 614, A37, \dodoi{10.1051/0004-6361/201732382}

\bibitem[{{Tortosa} {et~al.}(2021){Tortosa}, {Ricci}, {Tombesi}, {Ho}, {Du},
  {Inayoshi}, {Wang}, {Shangguan}, \& {Li}}]{Tortosa_2021}
{Tortosa}, A., {Ricci}, C., {Tombesi}, F., {et~al.} 2021, \mnras,
  \dodoi{10.1093/mnras/stab3152}

\bibitem[{{Turner} {et~al.}(2001){Turner}, {Abbey}, {Arnaud}, {Balasini},
  {Barbera}, {Belsole}, {Bennie}, {Bernard}, {Bignami}, {Boer}, {Briel},
  {Butler}, {Cara}, {Chabaud}, {Cole}, {Collura}, {Conte}, {Cros}, {Denby},
  {Dhez}, {Di Coco}, {Dowson}, {Ferrando}, {Ghizzardi}, {Gianotti}, {Goodall},
  {Gretton}, {Griffiths}, {Hainaut}, {Hochedez}, {Holland}, {Jourdain},
  {Kendziorra}, {Lagostina}, {Laine}, {La Palombara}, {Lortholary}, {Lumb},
  {Marty}, {Molendi}, {Pigot}, {Poindron}, {Pounds}, {Reeves}, {Reppin},
  {Rothenflug}, {Salvetat}, {Sauvageot}, {Schmitt}, {Sembay}, {Short},
  {Spragg}, {Stephen}, {Str{\"u}der}, {Tiengo}, {Trifoglio}, {Tr{\"u}mper},
  {Vercellone}, {Vigroux}, {Villa}, {Ward}, {Whitehead}, \&
  {Zonca}}]{Turner_2001}
{Turner}, M.~J.~L., {Abbey}, A., {Arnaud}, M., {et~al.} 2001, \aap, 365, L27,
  \dodoi{10.1051/0004-6361:20000087}

\bibitem[{{Wilkins} \& {Gallo}(2015)}]{Wilkins_2015}
{Wilkins}, D.~R., \& {Gallo}, L.~C. 2015, \mnras, 449, 129,
  \dodoi{10.1093/mnras/stv162}

\bibitem[{{Wu} {et~al.}(2020){Wu}, {Wang}, {Cai}, {Kang}, {Liu}, \&
  {Cai}}]{Wu_2020}
{Wu}, Y.-J., {Wang}, J.-X., {Cai}, Z.-Y., {et~al.} 2020, Science China Physics,
  Mechanics, and Astronomy, 63, 129512, \dodoi{10.1007/s11433-020-1611-7}

\bibitem[{{Zhang} {et~al.}(2018){Zhang}, {Wang}, \& {Zhu}}]{Zhangjx2018}
{Zhang}, J.-X., {Wang}, J.-X., \& {Zhu}, F.-F. 2018, \apj, 863, 71,
  \dodoi{10.3847/1538-4357/aacf92}

\end{thebibliography}
\bibliographystyle{aasjournal}

\end{document}